\documentclass[12pt]{article}
\usepackage{amsmath,amsfonts}
\usepackage[nosort]{cite}
\usepackage{graphicx}
\usepackage{feynmp}
\makeatletter\@addtoreset{equation}{section}\makeatother

\def\bC {\mathbb{C}}

\def\bP {\mathbb{P}}
\def\bR {\mathbb{R}}
\def\bZ {\mathbb{Z}}

\def\CP{\bC\bP}

\newcommand{\be}{\begin{equation}}
\newcommand{\ee}{\end{equation}}
\newcommand{\bea}{\begin{eqnarray}}
\newcommand{\eea}{\end{eqnarray}}

\newcommand{\vev}[1]{{\left< {#1} \right>}}

\newcommand{\eqn}[1]{(\ref{#1})}
\newcommand{\nn}{\nonumber}

\def\bsp{\be\begin{split}}

\def\G{\Gamma}

\def\a{\alpha}
\def\b{\beta}
\def\g{\gamma}

\def\d{\delta}
\def\e{\epsilon}
\def\m{\mu}
\def\n{\nu}
\def\s{\sigma}
\def\r{\rho}
\def\l{\lambda}
\def\t{\tau}
\def\o{\omega}

\def\vt{\vartheta}
\def\vp{\varphi}

\def\p{\partial}

\def\bR {\mathbb{R}}
\def\bZ {\mathbb{Z}}

\newcommand{\Tr}{{\rm Tr\,}}

\newcommand{\cL}{{\mathcal L}}
\newcommand{\cM}{{\mathcal M}}
\newcommand{\cN}{{\mathcal N}}
\newcommand{\cP}{{\mathcal P}}

\newcommand{\cS}{{\mathcal S}}

\newcommand{\hi}{{\hat i}}   
\newcommand{\hj}{{\hat j}}

\newcommand{\hA}{{\hat A}}   
\newcommand{\barC}{{\bar C}}

\newcommand{\btheta}{{\bar \theta}}

\newcommand{\slsh}[1]{{#1}\!\!\!\! / \, }
\newcommand{\pslash}{p\!\! /}
\newcommand{\kslash}{k\!\! /}

\renewcommand{\title}[1]{\vbox{\center\LARGE{#1}}\vspace{5mm}}
\renewcommand{\author}[1]{\vbox{\center#1}\vspace{5mm}}
\newcommand{\address}[1]{\vbox{\center\em#1}}
\newcommand{\email}[1]{\vbox{\center\tt#1}\vspace{5mm}}

\begin{document}
\begin{fmffile}{graphs}
\begin{titlepage}
\begin{center}
\vspace{5mm}
\hfill {\tt HU-EP-08/20}\\
\vspace{20mm}
\title{\bf\sc Wilson loops in 3-dimensional \mathversion{normal}$\cN=6$ 
\mathversion{normal} supersymmetric Chern-Simons Theory \\
and their string theory duals}
\vspace{10mm}
\author{\large Nadav Drukker, Jan Plefka and Donovan Young}

\address{Humboldt-Universit\"at zu Berlin, Institut f\"ur Physik,\\
Newtonstra{\ss}e 15, D-12489 Berlin, Germany}

\email{drukker,plefka,dyoung@physik.hu-berlin.de}

\end{center}
\vspace{10mm}

\abstract{
\noindent
We study Wilson loops in the three-dimensional $\cN=6$ supersymmetric 
Chern-Simons theory recently constructed by Aharony, Bergman, Jafferis 
and Maldacena, that is conjectured to be dual to type IIA string theory on $AdS_4\times
\CP^3$. We construct loop operators in the Chern-Simons theory
which preserve $1/6$ of the supercharges and calculate their expectation value up 
to 2-loop order at weak coupling. The expectation value at strong coupling is
found by constructing the string theory duals of these operators. For low dimensional
representations these are fundamental strings, for high dimensional representations
these are D2-branes and D6-branes. In support of this identification we demonstrate
that these string theory solutions match the symmetries, charges and the  
preserved supersymmetries of their Chern-Simons theory counterparts.
}

\vfill

\end{titlepage}

%{\addtolength{\parskip}{-1ex}\tableofcontents}

\section{Introduction}

This work focuses on supersymmetric Wilson loop operators in the 
three-dimensional Chern-Simons (CS) theory of Aharony, Bergman, Jafferis and Maldacena
\cite{Aharony:2008ug}. This 
theory is conjectured to represent the low-energy dynamics of $N$ 
coincident M2-branes at a $\bZ_k$ orbifold of the transverse $\bR^8$ 
space. This in turn has an alternative description
as a weakly coupled type IIA string theory on $AdS_4\times\CP^3$ 
(or more generally M-theory on $AdS_4\times S^7/\bZ_k$).

There are several reasons to focus on Wilson loop operators. They can be 
defined in any gauge theory and in the case of pure Chern-Simons theory, 
which is topological, they are the principal observables. While the theory of 
\cite{Aharony:2008ug} includes additional matter fields, Wilson loops 
are still very natural observables. Furthermore, these operators play an 
important role in the $AdS$/CFT correspondence \cite{Maldacena:1997re}, 
since they are dual to semiclassical 
strings in the dual supergravity background
\cite{Rey,Maldacena-wl}. Lastly, in the case of $\cN=4$ 
supersymmetric Yang-Mills (SYM) theory in four dimensions, 
the expectation value of the $1/2$ BPS circular Wilson loop is 
a non-trivial function of the 't Hooft coupling $\lambda$ and the 
rank of the gauge group $N$, yet it can be calculated exactly and 
matched with string theory 
\cite{Erickson:2000af,Drukker:2000rr,Pestun:2007rz}. 
It is therefore interesting to see if an 
analog observable exists in the 3-dimensional theory.

The supersymmetric Chern-Simons theory has two gauge groups 
of equal rank $N$ and opposite level $k$ and $-k$. In addition to the 
gauge fields there are bosonic and fermionic fields $C_I$ and 
$\psi_I$ respectively in the bi-fundamental $({\bf N},\bar{\bf N})$ representation 
of the gauge groups and their complex conjugates.

With two gauge groups and this matter content there are quite a few 
possibilities to construct gauge-invariant Wilson loop operators. One 
choice would simply be the standard Wilson loop operator in one of the 
gauge groups (with gauge field $A_\mu$ or $\hat A_\mu$)
\be
W=\frac{1}{N}\Tr\cP\exp\left(i\int A_\mu dx^\mu\right).
\label{bosonic-WL}
\ee
Our experience from $\cN=4$ SYM in 4-dimensions suggests that 
such a Wilson loop is not supersymmetric, which can be verified by 
a direct calculation.

In the four dimensional theory a supersymmetric Wilson loop couples 
also to an adjoint scalar field \cite{Maldacena-wl,Drukker:1999zq}. 
Here there are no adjoint fields, but we can use two bi-fundamental 
fields to construct a composite in the adjoint
\be
W=\frac{1}{N}\Tr\cP\exp \int\left(iA_\mu \dot x^\mu
+\frac{2\pi}{k}|\dot x|M^I_J C_I \bar C^J\right)ds\,.
\label{adjoint-WL}
\ee
$M^I_J$ is a matrix whose properties will be determined by supersymmetry. 
This is the Wilson loop operator we shall focus on.

With the appropriate choice of $M^I_J$, this Wilson loop will turn out to 
preserve $1/6$ of the supercharges (4 out of 24) when the path of the loop 
is a straight line or a circle. In the first case it has a trivial expectation value, 
but not in the case of the circle, where we calculate it to 2-loop order in the 
gauge theory and to leading order at strong coupling. For arbitrary shape 
it will not preserve global supersymmetry, but we still expect it to be the 
natural observable with a simple description in the string theory dual.

The next section studies the Wilson loop in the gauge theory and the following section 
does the same from the string-theory side.

In the course of this work we have learnt that some of our results were 
independently obtained by several other groups 
\cite{others,Berenstein:2008dc,Chen:2008bp}.

\section{Gauge theory construction}
\label{sec-1/6}

In this section we study the Wilson loop \eqn{adjoint-WL}. We classify
the conditions for it to be supersymmetric, 
derive the perturbative expression for this Wilson
loop and calculate it at two loop order in an expansion in the 't Hooft 
coupling $\lambda=N/k$.

\subsection{Supersymmetry}
\label{sec-1/6-susy}

The $\cN=6$ CS theory has 12 Poincar\'e supercharges
$(Q_{IJ})_\a=-(Q_{JI})_\a$, where $I,J=1,\ldots,4$, and the spinor
index takes the values $\a=1,2$. Along with the 12 superconformal supercharges
$S_{IJ}$, to be discussed below, these make up the 24 supersymmetries of the theory.
From \cite{Gaiotto:2008cg} we have the supersymmetry transformations of
the bosonic fields of the theory
\begin{equation}
\label{susygen}
\begin{aligned}
\delta C_K &=  (\theta^{IJ}\, Q_{IJ})\, C_K = \theta^{IJ}\, \varepsilon_{IJKL} \bar\psi^L,\\
\delta \barC^K&=(\theta^{IJ}\, Q_{IJ})\, \barC^K 
= \theta^{IJ}\left(\delta_I^K\psi_J - \delta_J^K \psi_I\right),\\
\delta A_\mu&= (\theta^{IJ}\, Q_{IJ})\, A_\mu = \frac{2\pi i}{ k}\,\theta^{IJ}\,
\sigma_\mu(C_{I} \psi_{J}-C_{J} \psi_{I}  + \varepsilon_{IJKL} \bar\psi^{K}\barC^{L}),\\
\delta \hat A_\mu&=(\theta^{IJ}\, Q_{IJ})\, \hat A_\mu = \frac{2\pi i}{ k} \,
\theta^{IJ}\,\sigma_\mu(\psi_{J}C_{I}-\psi_{I}C_{J} + \varepsilon_{IJKL} \barC^{L}\bar\psi^{K}).
\end{aligned}
\end{equation}
with the Poincar\'e supersymmetry parameter $(\theta^{IJ})^{\alpha}$. We note the complex conjugation
properties $\barC^K=(C_K)^\dagger$, $\bar \psi^K=(\psi_K)^\dagger$ 
and $(\theta^{IJ})^\dagger=\frac{1}{2} \varepsilon_{IJKL}\, \theta^{KL}$.

Let us then consider the supersymmetry variation of  the 
Wilson loop (\ref{adjoint-WL}) and demand that it vanishes for a suitable
choice of the $\theta^{IJ}$. One then finds the following condition
\be
\begin{aligned}
\delta W \sim & \,\,\, 
 \theta^{IJ}_\alpha\,  [-\dot x_\mu\, \sigma^\mu_{\alpha\beta}
\, \delta^P_I + |\dot x|\, \delta_{\alpha\beta}\, M^P_I\, ] \, C_p\, (\psi_J)_\beta\\
& +  \epsilon_{IJKL}\, \theta^{IJ}_\alpha\, [\dot x_\mu\, \sigma^\mu_{\alpha\beta}
\, \delta^K_P + |\dot x|\, \delta_{\alpha\beta}\, M^K_P\, ]\, (\bar\psi^L)_\beta \, \bar C^P
=0
\end{aligned}
\label{deltaW}
\ee
For a supersymmetric loop both terms in the above have to vanish seperately. 
Let us then consider a straight space-like Wilson line in the $1$ direction, 
i.e.~$x^\mu(s)=\delta^{1\mu}\, s$ and decompose the above equation with respect to the
projectors $P_\pm=\frac{1}{2}(1\pm\sigma^1)$.
We then find from \eqn{deltaW} the conditions
\begin{eqnarray}
& \theta^{IJ}_+\,  (-\delta^P_I+M^P_I)  +  \theta^{IJ}_-\,  (\delta^P_I+M^P_I)  
=0 \, ,\label{susycond1}&\\ 
&\varepsilon_{IJKL}\, \theta^{IJ}_+\, 
(\delta^K_P+M^K_P) + 
\varepsilon_{IJKL}\, \theta^{IJ}_-\, 
(-\delta^K_P+M^K_P) =0 \, ,&
\label{susycond2}
\end{eqnarray}
where $\theta^{IJ}_\pm=P_\pm \theta^{IJ}$. To analyze the possible solutions 
it is simplest to start with one specific supercharge, parameterized without loss 
of generality by a non-vanishing $\theta^{12}_+$. This choice implies
\be
M=\begin{pmatrix}
1&0&0&0\\
0&1&0&0\\
0&0&-1&0\\
0&0&0&-1
\end{pmatrix},
\qquad
M^I_JC_I\bar C^J=C_1\bar C^1+C_2\bar C^2-C_3\bar C^3-C_4\bar C^4\,.
\ee
It is simple to see that this choice of $M^I_J$ then 
allows for one more independent non-vanishing supercharge, 
parameterized by $\theta^{34}_-$.

This Wilson loop operator is therefore invariant under 
two out of the 12 Poincar\'e supersymmetries,
i.e. 1/6 of the super-Poincar\'e generators are preserved%
\footnote{Note that if the sign of the $\delta^K_P$ terms in \eqn{susycond2} 
was the opposite, the choice $M^I_J=\delta^I_J$ would preserve half 
the supercharges. Alas, this is not the case.}.

Let us now turn to the 12 super-conformal symmetries $(S^{IJ})_\alpha=-(S^{JI})_\alpha$.
The super-conformal transformations of the $\cN=6$ CS theory 
have been constructed recently in \cite{Schwarz}. For
the transformations of the bosonic fields the only change with respect to \eqn{susygen}
is the replacement $\theta^{IJ} \to x\cdot \sigma\, \eta^{IJ}$, while the 
super-conformal transformations of the fermionic fields receive an additional
contribution. This additional term, however, does not affect the variation of
the Wilson loop operator \eqn{adjoint-WL} and the super-conformal analogue of the
above  Wilson line 
analysis then results in the simple replacement of $\btheta^{IJ} \to {\bar \eta}^{IJ} s\,
 \sigma^1$ in \eqn{susycond1} and \eqn{susycond2}. 
 Hence, also two of the 12 super-conformal symmetries are
intact and we indeed find that the Wilson line operator \eqn{adjoint-WL} is 1/6 BPS.

This analysis is valid for an infinite straight line. Under a conformal 
transformation a line will be mapped to a circle, which will therefore 
posses the same number of supersymmetries. The conformal 
transformation mapping the line to the circle mixes the super-Poincar\'e 
and superconformal charges, hence the circular Wilson loop is 
invariant under a linear combination of $Q^{IJ}\pm S^{IJ}$.

These Wilson loops are invariant also under some bosonic 
symmetries, part of the $SO(4,1)\times SO(6)$ symmetry of the vacuum. 
There is an $SL(2,\bR)\times U(1)$ subgroup of the conformal group 
comprised, in the case of the line, 
of translations along the line $P_1$, dilation $D$, a special conformal 
transformation $K_1$ and a rotation around the line, $J$. These generators 
combine with the supercharges to form the supergroup 
$OSp(2|2)$ (with a non-compact $Sp(2)$). In addition there is an 
extra $SU(2)\times SU(2)$ subgroup of the $SO(6)$ R-symmetry group 
rotating $C_1\leftrightarrow C_2$ and $C_3\leftrightarrow C_4$ that 
leaves $M^I_J$, and hence the Wilson loop, invariant. 
The supercharges, being in the antisymmetric representation 
of the R-symmetry group are neutral under this extra bosonic symmetry.

Thus far we have discussed space-like Wilson loops. 
For a straight time-like Wilson loop 
we find similar conditions, only that the matrix $M$ will be imaginary. 
For a straight light-like line the scalar contribution to \eqn{adjoint-WL} 
vanishes, but the loop is still supersymmetric. In this case it is invariant 
under half of the super-Poincar\'e charges and all the super-conformal 
ones. The fact that the scalar coupling is real for a space-like curve, 
imaginary for a time-like one and vanishes for a light-like curve 
is familiar from $\cN=4$ SYM in four 
dimensions \cite{Drukker:1999zq}. 

Given a choice of supercharges it is an interesting
question to ask what is the most general loop preserving it. We saw
that the basic Wilson loop \eqn{adjoint-WL} with the geometry of a
line or a circle preserves four real supercharges. Under this choice of
supercharges the matrix $M^I_J$ was fixed, as was the value of 
$\dot x_\mu$. So the loop is restricted to be a line in a fixed
direction. Parallel lines will preserve the same super-Poincar\'e charges, 
but different superconformal ones.

Thus the choice of four supercharges completely fixes the geometry of the 
loop. However, this does not mean that 
there is only a unique Wilson loop preserving these supercharges, there are 
different ones with the same geometry but in different representations of the 
gauge groups.
 
In \eqn{adjoint-WL} we chose one of the gauge groups, but a similar operator 
exists also in the other group. In that case instead of $C_I\bar C^J$ the scalar bilinear 
will be of the opposite order $\bar C^J C_I$. More generally, 
we can take the Wilson loop to be in any representation of each of 
the gauge groups, so the most general Wilson loop will be characterized by 
a pair of Young tableau for the representations $R$ and $\hat R$
\be
W^\pm_{R\hat R}=\frac{1}{2}\left[\text{Tr}_{R}\cP e^{\int\left(iA_\mu \dot x^\mu
+\frac{2\pi}{k}|\dot x|M^I_J C_I \bar C^J\right)ds}
\pm\widehat{\text{Tr}}_{\hat R}\cP e^{\int\left(i\hat A_\mu \dot x^\mu
+\frac{2\pi}{k}|\dot x|\bar M^I_J \bar C^J C_I\right)ds}\right].
\label{double-WL}
\ee

This in fact over-counts the number of Wilson loops. Recall that in 
Chern-Simons theory there are 't Hooft vertices which are in the $k$'th symmetric 
representation \cite{'tHooft:1977hy,Moore:1989yh}. These are important to 
create some of the local gauge invariant states in the theory \cite{Aharony:2008ug}, 
but they also affect the Wilson loops. Since they can be added freely, they 
essentially identify representations which are related to each-other by 
multiplication by the $k$'th symmetric representation. Thus they reduce the 
number of distinguished Wilson loop observables to be those given by 
Young tableau with fewer than $k$ columns.

Furthermore, it could also be quite difficult to find all of those different Wilson loops in 
the supergravity limit. In similar cases (like in orbifolds of $\cN=4$ of SYM) only 
the Wilson loops that are symmetric under interchange of the gauge groups 
have a known simple description. 
In this theory the most natural operator of the type \eqn{double-WL} 
is the one that is symmetric under the 
exchange of the two gauge groups, while exchanging also the representation 
with its conjugate (since the matter is in the fundamental - anti-fundamental). 

We expect therefore our string theory solutions presented in 
Section~\ref{sec-string} to correspond to this linear combination of 
Wilson loops in the two gauge groups. The leading planar contribution will be 
a single string, dual to a single-trace Wilson loop (or a multiply wrapped 
Wilson loop). For very large representations the planar approximation 
breaks down and the fundamental string should be supplanted by D-branes.

\subsection{Perturbative calculation}
\label{sec-pert}

Let us now turn to the perturbative evaluation in $\lambda=N/k$ of 
the 1/6 BPS Wilson loop \eqn{adjoint-WL}
for circular and straight line contours. We shall work in Euclidean space. 
At leading order in $\lambda$ the only possible contribution is from a tree-level 
gluon exchange which is identical to that in pure CS theory. The result is 
rather subtle and depends on the ``framing'' of the Wilson loop, which 
is extra information needed to define it beyond the path of the loop 
(c.f. \cite{Witten:1988hf,Guadagnini:1989am,Alvarez:1991sx}). 
We will take a slightly naive approach; since the gluon propagator is 
proportional to the antisymmetric epsilon tensor, it vanishes for all loops 
lying in a plane. This corresponds to zero framing. A possible additional subtlety
arises from the self contraction of the two scalars fields at leading order. We take them
to be defined as normal ordered. Hence there is no contribution at leading order in $\lambda$.

Expanding $W$ to second order we need the one-loop corrected Feynman gauge gluon propagator
and the bare scalar propagator calculated in Appendix~\ref{appA} (see also
\cite{Gaiotto:2007qi})
\begin{align}
 \langle
A_\mu(x)_{ij} A_\nu(y)_{kl}\rangle &= {\delta}_{ik}\d_{jl} \frac{1}{k}
\left[-i\frac{\varepsilon_{\mu\nu\rho}(x-y)^\rho}{2|x-y|^3}
+\frac{N}{k} \left(\frac{\delta_{\mu\nu}}{|x-y|^2 }-\partial_\mu
\partial_\nu \ln|x-y|\right)\right] \, ,
\nonumber\\
\langle (C_I)_{i\hi}(x)\, (\bar C^J)_{\hj j}(y) \rangle &=  \delta^J_I\,
\delta_{ij}\, \delta_{\hi \hj}\,
\frac{1}{4\pi |x-y|} \, .
\end{align}
\begin{figure}[t]
\begin{center}
(a)\,
\raisebox{-0.8cm}{\begin{fmfgraph}(20,20)
 \fmfleft{l}\fmfright{r}\fmftop{t}\fmfbottom{b}
\fmf{plain,left,width=2}{l,r,l}\fmf{boson}{l,v1}\fmf{boson}{v1,r}\fmfblob{0.25w}{v1}
\end{fmfgraph}}
\qquad
(b)\,
\raisebox{-0.8cm}{
\begin{fmfgraph}(20,20)
 \fmfleft{l}\fmfright{r}\fmftop{t}\fmfbottom{b}
\fmf{plain,left,width=2}{l,r,l}\fmf{dbl_plain}{l,r}
\end{fmfgraph}}
\qquad
(c)\,
\raisebox{-0.8cm}{
\begin{fmfgraph}(20,20)
 \fmfleft{l}\fmfright{r}\fmftop{t}\fmfbottom{b}
\fmf{plain,left,width=2}{l,r,l}\fmf{dbl_plain}{l,v}\fmf{boson}{v,r}\fmfdot{v}
\end{fmfgraph}}
\qquad
(d)\,
\raisebox{-0.8cm}{
\begin{fmfgraph}(20,20)
\fmfsurround{l,t,r}\fmfleft{ll}
\fmf{plain,left,width=2}{ll,l,ll}
\fmf{boson}{l,v,r}\fmf{boson}{t,v}\fmfdot{v}
\end{fmfgraph}}
\end{center}
\caption{The two-loop Feynman diagrams contributing to a circular $\langle W\rangle$. 
The bold circular line represents the Wilson loop contour, whereas
wiggly lines denote gluon and straight lines scalar propagators.}
\label{fig:2loop_graphs}
\end{figure}
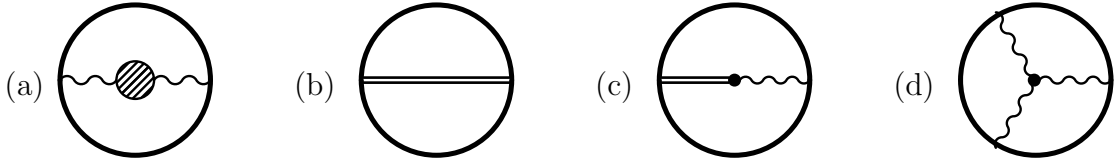
At this two-loop order one finds that in the loop-to-loop propagator the 
propagator of the composite scalar $M^I_J\, \barC^J C_I$, diagram (b),
combines with the one-loop piece of the gauge field propagator, diagram (a), 
to give 
\be
{\cal D}[x_1(\tau_1), x_2(\tau_2)] \equiv -\frac{N^3}{k^2}\, 
\left [\,\frac{\dot x_1\cdot \dot x_2 - |\dot x_1| |\dot x_2| }{(x_1-x_2)^2}
-\partial_{\tau_1} \partial_{\tau_2} \ln |x_1-x_2|\,\right ] \, .
\label{combinedprop}
\ee
We would like to point out a subtlety in the last term, which being a total 
derivative can be removed by a gauge transformation -- albeit a singular one.
Depending on the regularization it may lead to divergences along the loop, 
as we do not expect divergencies for the supersymmetric Wilson loop we conclude 
that it should be dropped.
Also note that the scalar contribution is insensitive to the choice of signs in the
$\pm 1$ entries of the diagonal $M^I_J$ as these come in squares. One sees that
for a straight line this yields a vanishing effective propagator,
while for the circle it gives a constant propagator ${\cal D}=N^3/(2k^2)$ 
somewhat  similar to the situation in four dimensional ${\cal N}=4$ super Yang-Mills. 
Thus this contribution gives at ${\cal O}(k^{-2})$
\be
\frac{1}{N}\frac{1}{2!}\oint d\t_1 \oint d\t_2 \, \frac{N^3}{2k^2}
=  \frac{\pi^2 N^2}{k^2}\, .
\ee

There are two other diagrams contributing at ${\cal O}(k^{-2})$. The
diagram (c) is the interaction between a scalar bilinear $C_I\,\barC^J$ 
and a gauge field
\begin{equation}\label{BBA}
\begin{split}
\Biggl< \frac{2}{N} \Tr \oint d\tau_1\,d\tau_2\, i \dot x_1^\mu |\dot x_2|
\frac{M^I_J}{k} A_\mu (x_1) 
 C_I \barC^J(x_2) 
\int d^3 w \, \Tr \left( i \partial_\rho  C_K A_\rho \barC^K 
-i \partial_\rho \barC^K C_K  A_\rho \right) \Biggr>\\
\propto  \oint d\tau_1\,d\tau_2\,\int d^3 w\, 
\varepsilon_{\mu \nu \rho} \,\dot x_1^\mu \,|\dot x_2| \,(x_1-w)^\rho
\frac{1}{|x_1-w|^3|x_2-w|} \frac{\partial}{\partial w^\nu}
\frac{1}{|x_2-w|}=0\,.
\end{split}
\end{equation}
It is zero because the integrand is odd in the third component of $w$,
i.e.~the component orthogonal to the plane of the circular loop.

 The remaining
diagram (d) is an interaction of three gauge fields through the
Chern-Simons interaction. 
This graph appears also in pure
Chern-Simons theory and its value depends only on the topology of the
loop. The circle is an ``unknot'', for which the result is
$-N^2\pi^2/(6k^2)$ \cite{Guadagnini:1989am}.

Putting together the ${\cal O}(k^{-2})$ contributions we find
\be
\langle W \rangle = 1 + \frac{\pi^2 N^2}{k^2} -
\frac{\pi^2 N^2}{6 k^2} + {\cal O}(k^{-3})\,,
\ee
where we have separated the $O(k^{-2})$ contribution into two terms, one
from the combined gauge-field and scalar exchange and the second, the
topological contribution identical to pure Chern-Simons.

So far we discussed the Wilson loop in one of the two $U(N)$ factors, 
but it makes sense to consider the linear combination of the operators in 
the two groups \eqn{double-WL}. 
In particular we expect the string theory duals to be 
symmetric under the exchange of the two groups. 
As mentioned before, one would be lead to take the Wilson loop in the 
conjugate representation, which can be simply expressed as the usual 
Wilson loop with an overall sign reversed.

The perturbative calculation for the second gauge group is identical to the 
first up to some sign changes. The sign of the level $k$ is reversed, which 
will change the signs of the propagators and the interaction vertex. The total 
number of them in all of the graphs of order $\lambda^2$ is always 
even, so that will not create any change. But the overall sign in the Wilson 
loop is also reversed which will affect 
the signs of the graphs where the loop was expanded to odd-order. In our 
case there is only one such graph, Fig.~\ref{fig:2loop_graphs}d.
This is the graph that gave the pure CS contribution.

Therefore at the 2-loop order if we consider the two possible linear 
combinations of the loops in the two gauge groups in the fundamental and 
anti-fundamental representations, the sum of the two 
will not include the CS term and the difference will include only the 
CS contribution%
\footnote{The possibility for such a cancelation was first observed in 
\cite{Chen:2008bp}, though for a somewhat different construction. 
See also \cite{Rey:2008bh}}.

For the $1/2$ BPS circular Wilson loop in $\cN=4$ in four-dimensions the gauge 
field and scalar propagators combined to a constant, similar to 
what we have found here at order $\lambda^2$. In four dimensions the 
interactions also cancel and the full answer is given by summing over 
the free constant propagators, i.e. a zero-dimensional Gaussian matrix 
model \cite{Erickson:2000af,Drukker:2000rr}. 
In that case the result in the planar approximation can be expressed 
in terms of a Bessel function
\be
\langle W_{\cN=4}\rangle_\text{planar}
\sim\frac{2}{\sqrt\lambda}\, I_1(\sqrt\lambda)
=
\begin{cases}
1+ \frac{1}{8}\, \lambda + \ldots & \text{for} \quad \lambda\ll 1 \cr
e^{\sqrt{\lambda}} & \text{for} \quad  \lambda\gg 1 \cr
\end{cases}
\label{mm}
\ee

\begin{figure}[t]
\begin{center}
\raisebox{-0.8cm}{
\begin{fmfgraph}(20,20)
\fmfsurround{l,t,r}\fmfleft{ll}
\fmf{plain,left,width=2}{ll,l,ll}
\fmf{plain}{l,r}\fmf{plain}{l,t}\fmf{plain}{t,r}
\end{fmfgraph}}
\qquad
\raisebox{-0.8cm}{
\begin{fmfgraph}(20,20)
\fmfsurround{l,t,r,u}\fmfleft{ll}
\fmf{plain,left,width=2}{ll,l,ll}
\fmf{plain}{u,r}\fmf{plain}{r,t}\fmf{plain}{t,l}\fmf{plain}{l,u}
\end{fmfgraph}}
\end{center}
\caption{Some examples of higher level N-gon tree graphs.}
\label{box}
\end{figure}
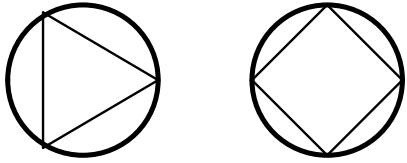

In the case at hand it is clear that interactions do contribute. For one, 
the constant propagator ${\cal D}=N^3/(2k^2)$ emerged from the sum 
of the one-loop corrected gluon self-energy and the tree-level scalar 
exchange \eqn{combinedprop}. Furthermore independent of that we also found the 
result of pure Chern-Simons. Another novel 
feature is that the tree-level graphs do not only have ladder structure. 
Rather, there will be in general tree level graphs of N-gon topology due to the biscalar coupling in the loop, see figure~\ref{box}.
%

% Yet we would like to point out that the result of the perturbative 
% calculation excluding the pure-CS contribution matches with the 
% first term in the matrix model result \eqn{mm} as does the 
% string theory calculation in the next section under the replacement
% \be
% \sqrt\lambda\to2\pi h(\lambda)\,,\qquad
% h(\lambda)
% =
% \begin{cases}
% \lambda + O(\lambda^2) & \text{for} \quad \lambda\ll 1 \cr
% \sqrt{\lambda/2} + O(1) & \text{for} \quad  \lambda\gg 1 \cr
% \end{cases}
% \label{hguess}
% \ee
% This definition of $h(\lambda)$ exactly matches with the novel function 
% appearing in the dispersion relation of magnons 
% \cite{Nishioka:2008gz,Gaiotto:2008cg,Grignani:2008is}, 
% which we find to be an interesting 
% coincidence.

Let us also note that in the case of $\cN=4$ SYM in four dimensions there 
are other BPS Wilson loops preserving fewer supercharges whose perturbative 
expansions are rather complicated and do include interacting graphs. Still 
there is some evidence that they are given by the same answer as the circular 
Wilson loop \eqn{mm}, only with a rescaled coupling
\cite{D06,Semenoff:2006am,Bassetto:2008yf,Young:2008ed}.

%We would like to stress that beyond a matching of the numerical coefficients 
%of the effective gluon and bi-scalar propagator and that of the string tension, 
%the expression \eqn{hguess} is unmotivated. 
While there is no strong evidence for a simple cancelation, we still 
find it conceivable that the $1/6$ BPS circular Wilson loop will also have 
an exact perturbative result that can be resummed to all orders, like the 
supersymmetric Wilson loops in four-dimensions. If indeed so, then 
to get a match with the string theory result in the next section
the coupling in the analogeous matrix model result \eqn{mm} 
would certainly have to be renormalized in some way.

\section{String theory description}
\label{sec-string}

The three-dimensional $\cN=6$ CS theory is conjectured to be dual to 
M-theory on $AdS_4\times S^7/\bZ_k$. To understand the action of the 
$\bZ_k$ orbifold, one should write $S^7$ as a circle fibration over 
complex projective space $\CP^3$, where the orbifold acts on the fiber 
(see \eqn{S7-metric} below). 
For large $k$ the radius of this ``M-theory circle'' becomes small, so the 
theory can be described in terms of type IIA string theory 
on $AdS_4\times \CP^3$ with string-frame metric
\be
ds^2=\frac{R^3}{4k}\left(ds^2_{AdS_4}+4ds^2_{\CP^3}\right)\,.
\label{metric}
\ee
We choose in this paper to work in the string 
theory picture, but all the solutions we describe below should also have an 
uplift to the full M-theory.

For the $AdS_4$ part we may use the global Lorentzian metric
\be
ds_{AdS_4}^2=-\cosh^2\rho\,dt^2+d\rho^2
+\sinh^2\rho\big(d\theta^2+\sin^2\theta\,d\psi^2\big)\,.
\label{AdS-metric}
\ee
or as foliated by $AdS_2$ slices
\be
\begin{gathered}
ds_{AdS_4}^2=du^2+\cosh^2u\,ds_{AdS_2}^2+\sinh^2u\,d\phi^2\,.
\\
ds_{AdS_2}^2=
\begin{cases}
d\rho^2-\cosh^2\rho\,dt^2\,,\qquad\text{appropriate for a time-like line,}\\
d\rho^2+\sinh^2\rho\,d\psi^2\,,\qquad\text{appropriate for a space-like circular loop.}
\end{cases}
\label{AdS-metric-2}
\end{gathered}
\ee
The metric on $\CP^3$ can be written in terms of four complex projective
coordinates $z_i$ as
\be
ds_{\CP^3}^2=\frac{1}{\rho^2}\sum_{i=1}^4 dz_i\,d\bar z_i
-\frac{1}{\rho^4}\bigg|\sum_{i=1}^4 z_i\,d\bar z_i\bigg|^2\,,\qquad
\rho^2=\sum_{i=1}^4|z_i|^2\,.
\ee

In the following we choose a specific representations in terms of angular 
coordinates (used also in \cite{Cvetic:2000yp,Nishioka:2008gz}). 
We start by parametrizing $S^7\subset\bC^4$ as
\be
\begin{aligned}
z_1&=\cos\frac{\alpha}{2}\cos\frac{\vartheta_1}{2}\,e^{i(2\varphi_1+\chi+\zeta)/4}\,,\\
z_2&=\cos\frac{\alpha}{2}\sin\frac{\vartheta_1}{2}\,e^{i(-2\varphi_1+\chi+\zeta)/4}\,,\\
z_3&=\sin\frac{\alpha}{2}\cos\frac{\vartheta_2}{2}\,e^{i(2\varphi_2-\chi+\zeta)/4}\,,\\
z_4&=\sin\frac{\alpha}{2}\sin\frac{\vartheta_2}{2}\,e^{i(-2\varphi_2-\chi+\zeta)/4}\,,\\
\end{aligned}
\label{zi}
\ee
The metric on $S^7$ is then given by
\begin{align}
ds^2_{S^7}=&\frac{1}{4}\Bigg[
d\alpha^2
+\cos^2\frac{\alpha}{2}(d\vartheta_1^2+\sin^2\vartheta_1^2\,d\varphi_1^2)
+\sin^2\frac{\alpha}{2}(d\vartheta_2^2+\sin^2\vartheta_2^2\,d\varphi_2^2)
\nn\\&\hskip1cm
+\sin^2\frac{\alpha}{2}\cos^2\frac{\alpha}{2}
(d\chi+\cos\vartheta_1\,d\varphi_1-\cos\vartheta_2\,d\varphi_2)^2
+\frac{1}{4}(d\zeta+A)^2\, \Bigg]\,,
\label{S7-metric}\\
A=&\cos\alpha\,d\chi+2\cos^2\frac{\alpha}{2}\cos\vartheta_1\,d\varphi_1
+2\sin^2\frac{\alpha}{2}\cos\vartheta_2\,d\varphi_2\,.
\label{S7-form}
\end{align}
The angle $\zeta$ appears only in the last term and if we drop it 
we end up with the metric on $\CP^3$
\be
\begin{aligned}
ds^2_{\CP^3}=\frac{1}{4}\bigg[&
d\alpha^2
+\cos^2\frac{\alpha}{2}(d\vartheta_1^2+\sin^2\vartheta_1^2\,d\varphi_1^2)
+\sin^2\frac{\alpha}{2}(d\vartheta_2^2+\sin^2\vartheta_2^2\,d\varphi_2^2)
\\&
+\sin^2\frac{\alpha}{2}\cos^2\frac{\alpha}{2}
(d\chi+\cos\vartheta_1\,d\varphi_1-\cos\vartheta_2\,d\varphi_2)^2\bigg].
\end{aligned}
\label{CP3-metric}
\ee
The ranges of the angles are
$0\leq\alpha,\vartheta_1,\vartheta_2\leq\pi$, 
$0\leq\varphi_1,\varphi_2\leq2\pi$ and $0\leq\chi\leq4\pi$.

In addition to the metric, the supergravity background has the
dilaton, and the 2-form and 4-form field strengths from the
Ramond-Ramond (RR) sector
\be
e^{2\Phi}=\frac{R^3}{k^3}\,,
\qquad
F_4=\frac{3}{8}\,R^3\,d\Omega_{AdS_4}\,,
\qquad
F_2=\frac{k}{4}\,dA\,.
\label{field-strengths}
\ee
Here $d\Omega_{AdS_4}$ is the volume form on $AdS_4$ and $F_2$ is
proportional to the K\"ahler form on $\CP^3$.

To write down the general D-brane action in this background one also 
needs the potentials for these forms. 
The one-form potential is, up to gauge transformations
\be
C_1=\frac{k}{4}\,A\,,
\label{C1}
\ee
With $A$ defined in \eqn{S7-form}.

$C_3$, the three-form potential for $F_4$ will actually not play a role in our current 
calculations, but we write it down for completeness. 
The forms are defined in principle only up to a gauge choice, but since $C_3$ 
involves the non-compact directions and it may couple to branes that 
approach the boundary of space, one should impose a proper
asymptotic behavior on it. It seems like the analog of choosing 
Fefferman-Graham coordinates \cite{Fefferman} is to take the 3-form to
not have any component in the $du$ direction in the coordinate systems in 
\eqn{AdS-metric-2}. Such a prescription indeed 
gave the correct result in $\cN=4$ SYM in four dimensions 
\cite{Drukker:2005kx}%
\footnote{See a more detailed discussion in \cite{Drukker:2008wr}.}.
We therefore have for the three-form potential 
\be
C_3=\frac{1}{8}R^3\cosh^3u\times
\begin{cases}
\cosh\rho\,dt\wedge d\rho\wedge d\phi\,,
\qquad\text{appropriate for a time-like line}\\
\sinh\rho\,d\psi\wedge d\rho\wedge d\phi\,,
\qquad\text{appropriate for a space-like circular loop.}
\end{cases}
\label{C3}
\ee

The dual of $F_4$ is proportional to the volume form on $\CP^3$
\be
F_6=\star F_4=
\frac{3R^6}{2^8k}\sin^3\alpha\sin\vartheta_1\sin\vartheta_2\ 
d\alpha\wedge d\vartheta_1\wedge d\vartheta_2\wedge d\chi
\wedge d\varphi_1\wedge d\varphi_2\,.
\ee
The five-form potential for $F_6$ can then be written as
\be
C_5=-\frac{R^6}{2^8k}\,(\sin^2\alpha\cos\alpha+2\cos\alpha-2)
\sin\vartheta_1\sin\vartheta_2\ d\vartheta_1\wedge d\vartheta_2
\wedge d\chi\wedge d\varphi_1\wedge d\varphi_2\,.
\label{C5}
\ee
Here we chose a gauge that is regular at $\alpha=0$. Reversing the sign 
on the $-2$ term in the parentheses gives the gauge that is regular at 
$\alpha=\pi$.

The relation between the parameters of the string background and of
the field theory are (for $\alpha'=1$)
\be
\frac{R^3}{4k}=\pi\sqrt{\frac{2N}{k}}=\pi\sqrt{2\lambda}\,.
\label{radius-1}
\ee

\subsection{Fundamental string}
\label{sec-F1}

In the strong coupling description of $\cN=4$ SYM in terms of type IIB 
string theory on $AdS_5\times S^5$, a Wilson
loop in the fundamental representation is given by a 
fundamental string ending along the path of the loop on the boundary
of space. We expect this property to extend from $\cN=4$ in four
dimensions to our Wilson loops in the 3-dimensional CS theory.

In $\cN=4$ SYM the natural Wilson loop carries an $SO(6)$ vector 
index, representing its position on $S^5$, the analog for $\CP^3$ would 
be the fundamental representation of $SU(4)$, though we saw that the
$1/6$ BPS Wilson loop couples to two scalars, one in the $\bf4$ representation 
and the other in the $\bar{\bf4}$ with a matrix $M^I_J$. This matrix breaks 
$SU(4)\to SU(2)\times SU(2)$, so the string theory dual should not 
be localized at a point on $\CP^3$ (which would break 
$SU(4)\to U(3)$) but rather smeared along a $\CP^1$.

Still, if the scalar couplings are constant along the loop, we can forget 
about the $\CP^3$ part of the $\sigma$-model and focus on $AdS_4$. 
Any known string solution found in $AdS_5$ which can be embedded within 
an $AdS_4$ subspace is immediately a solution for this theory. So, 
many results that were derived for Wilson loops in $\cN=4$ SYM 
are valid also for our Wilson loops in $\cN=6$ CS.
For example the expressions for the anti-parallel lines 
(``quark - anti-quark potential'' \cite{Rey,Maldacena-wl}) and for the
light-like cusp \cite{Kruczenski:2002fb}
are exactly the same in the planar limit up to the change
$\lambda_{\cN=4}\to2\pi^2\lambda_{CS}$. 
A similar result for the cusp anomalous dimension 
was obtained from rotating strings in \cite{Aharony:2008ug}. 

In this paper we focus on supersymmetric configurations of straight 
lines or circles. The analog of the straight line on the $S^2\times\bR$ 
boundary of global $AdS_4$ \eqn{AdS-metric} is a pair of anti-parallel lines at 
antipodal points on $S^2$ (or in the coordinate system \eqn{AdS-metric-2} one 
sets $u=0$). The string solution describing them is 
an $AdS_2$ subspace spanned by the coordinates $\rho$ and $t$. 
After subtracting a divergence, the resulting action vanishes, meaning 
that the expectation value of the Wilson loop is unity.

To describe the circular Wilson loop one could use the Poincar\'e patch metric, 
as was done in \cite{Berenstein:IJ,Drukker:1999zq}, or use global 
$AdS_4$ and for simplicity take the circle to wrap a big circle on 
$S^2$, i.e. $\theta=\pi/2$ at constant time $t$ (or $u=0$ in the metric 
\eqn{AdS-metric-2}). The string solution will now be a Euclidean $AdS_2$ 
section spanned by $\rho$ and $\psi$. The action is proportional
to the area
\be
\cS_\text{string, cl.}=\frac{R^3}{8\pi k}
\int_0^{2\pi}d\psi\int_0^{\rho_0} d\rho\,\sinh\rho
=\pi\sqrt{2\lambda}\left(\cosh\rho_0-1\right)\,.
\ee
Here $\rho_0$ is a cutoff near the boundary of $AdS_2$ (which is also
at the boundary of $AdS_4$) and we expect the divergent term to be
removed by a boundary term as in \cite{Drukker:1999zq}. Using the standard
$AdS$/CFT dictionary we derive
\be
\vev{W}_\text{string}\sim e^{\pi\sqrt{2\lambda}}\,.
\ee

As mentioned before, this string would not be localized on $\CP^3$, but has to 
be smeared on a $\CP^1$. This can be the sphere parameterized by $\vartheta_1$ 
and $\varphi_1$ at $\alpha=0$ in the coordinate system \eqn{CP3-metric}. 
As mentioned before, in the string theory picture there isn't a simple way of 
distinguishing between the two gauge groups. We expect this string (as 
well as the D-branes discussed below) to correspond to a linear combination 
of Wilson loops which is symmetric under the exchange of the two gauge groups. 
Note that this is also the combination in the gauge theory where the pure 
Chern-Simons term at order $\lambda^2$ dropped out.

The uplift of this string solution to M-theory is straight forward.

\subsection{D2-brane}
\label{sec-D2}

In $\cN=4$ SYM in four dimensions a Wilson loop in a low dimensional 
representation is well represented at strong coupling by a free 
string in $AdS_5\times S^5$. For representations 
of dimension of order $N$ a better description is in terms of D3-branes 
(the symmetric representation) or D5-branes (antisymmetric)
\cite{Rey,Drukker:2005kx,
Yamaguchi:2006tq,Gomis:2006sb,Okuyama:2006jc,Hartnoll:2006is,Gomis:2006im}.
This is the Wilson loop version of a giant-graviton 
\cite{McGreevy:2000cw,Grisaru:2000zn,Hashimoto:2000zp}, 
sometimes also referred to as ``giant Wilson loop.''
For even higher dimensional representations the branes back-react on the 
geometry and one instead finds ``bubbling geometries'' 
\cite{Lin:2004nb,Yamaguchi:2006te,Lunin:2006xr,D'Hoker:2007fq,Okuda:2008px,Gomis:2008qa}.

In this subsection we present a D2-brane solution that is a possible candidate for 
a dual of Wilson loops. In the next subsection we present a D6-brane solution. 
In support of the identification with Wilson loop operators are their symmetries, 
their charges, classical action, and the supercharges they preserve.

Since the Wilson loop has an $SL(2,\bR)$ symmetry we expect the 
D2-brane to have an $AdS_2$ factor, which will be inside $AdS_4$. 
The third world-volume direction will be compact --- a circle. 
We therefore take as world-volume coordinates $\rho$, $t$ from 
\eqn{AdS-metric} (or alternatively $\rho$ and $t$, or $\rho$ and $\psi$ 
from \eqn{AdS-metric-2} with $u=0$) and a third world-volume coordinate
$\t$ of period $2\pi$.

We have found a few different solutions to the equations of motion of the 
D2-brane with this circle made of the $\phi$ circle at non-zero $u$ in 
$AdS_4$ \eqn{AdS-metric-2} and/or a circle inside $\CP^3$ 
similar to those of \cite{Lunin:2007ab}. While 
the experience from $AdS_5\times S^5$ might lead one to suspect that 
the dual of the Wilson loop in the symmetric representation should have 
the circle inside $AdS_4$, these solutions have 
a different gauge-theory interpretation \cite{next}. 
The most likely candidate for a dual of the Wilson loop has the circle 
inside $\CP^3$.

Since our Wilson loops have an $SU(2)\times SU(2)$ symmetry which 
acts by rotating $z_1$ into $z_2$ and $z_3$ into $z_4$, it is natural to 
take the circle to be in the $\chi$ direction, i.e. $\chi=-2\tau$ 
(recall that $\chi$ has period $4\pi$ and the choice of sign seems to 
be dictated by suspersymmetry). We would still need to set its 
location in terms of the other angles $\vartheta_1$, $\varphi_1$, 
$\vartheta_2$, $\varphi_2$ and $\alpha$. For now we take all of them to 
be constants, which seems to be a consistent ansatz. At the end, in order 
to restore the $SU(2)\times SU(2)$ symmetry (and the correct supersymmetry) 
we will smear the brane over the $\vartheta_1$, $\vartheta_2$, $\varphi_1$ 
and  $\varphi_2$ directions.

The action includes the Dirac-Born-Infeld (DBI) piece and the
Wess-Zumino (WZ) coupling
\be
\cS_\text{D2}=T_\text{D2}\int e^{-\Phi}\sqrt{\det(g+2\pi\alpha'F)}
+T_\text{D2}\int\Big[P[C_3]+2\pi i\alpha'P[C_1]\wedge F\Big].
\label{D2-action}
\ee
Here $g$ is the induced metric and $F$ is the intrinsic field strength
 on the world-volume. To describe a Wilson loop, which carries electric charge
the component $F_{t \rho }=E\cosh\rho$ will be non-zero, in the Lorentzian 
case. For the dual of the space-like circular loop, which is the case we 
work out in detail, it will instead be $F_{\psi\rho}=E\sinh\rho$.
Being that it represents an electric 
field and that the signature is Euclidean, it is imaginary. $P[C_3]$
is the pullback of the RR three-form potential, which vanishes on our 
configuration and $P[C_1]$ is the pullback of
the one-form. The last term comes with an $i$ again due to the fact
that we are in Euclidean signature.

After fixing all the other angles, the angles $\alpha$ and $\chi/2$ parameterize an 
$S^2$ of radius $1/2$. 
The field-strength $F_2$ in \eqn{field-strengths} is that of $k/2$ Dirac 
monopoles, but the one-form \eqn{C1} with $A$ as in \eqn{S7-form} 
is singular at both $\alpha=0$ and $\alpha=\pi$. 
Instead we take
\be
C_1=\frac{k}{4}(\cos\alpha-1)\,d\chi\,, 
\ee
which is regular at $\alpha=0$. The same expression with $(\cos\alpha+1)$ 
will be regular at $\alpha=\pi$.

Plugging our ansatz in we find
\be
\cS_\text{D2}=\frac{T_\text{D2}R^3}{8}\int d\rho\,d\psi\,d\tau\,\sinh\rho
\left[\sin\alpha\sqrt{1+\beta^2E^2}
-i\beta E(\cos\alpha-1)\right]\,,
\ee
with $\beta=8\pi k/R^3=\sqrt{2/\lambda}$ (setting $\alpha'=1$).
and note that we are using conventions where the D2-brane tension is
$T_\text{D2}=1/4\pi^2$.

The equation of motion for $\alpha$ allows it to be an arbitrary constant but 
gives the relation
\be
i\beta E=-\cos\alpha\,.
\ee
The gauge field is a cyclic variable and the flux through the brane is
proportional to the conjugate momentum
\be
p=-4\pi i\,\frac{\delta\cL}{\delta F}
=\frac{k}{2}\,.
\label{p-D2}
\ee

Now we wish to evaluate the action on this classical solution. 
As is explained in \cite{Drukker:2005kx}, the action as it stands does 
not give the correct classical value, since it is a functional of the
electric field and one should take a Legendre transform to replace $E$ 
by $p$. The result is
\be
\cS_\text{L.T, classical}=\cS_\text{clasical}-pE=\frac{R^3}{8}\int d\rho\,\sinh\rho
=\frac{k}{2}\,\pi\sqrt{2\lambda}(\cosh\rho_0-1)
\ee
Once we remove the divergence from large $\rho$, we see that this solution 
agrees with that of $k/2$ fundamental strings.

The charge and action agree exactly with that of $k/2$ fundamental strings, 
while the angle $\alpha$ is completely arbitrary. To see if there 
are solutions with $|p|<k/2$ it is useful to consider the Legendre transform 
before solving the equations of motion. The action in terms of $p$ is
\be
\cS_\text{L.T.}=\cS_\text{D2}-pE
=\frac{T_\text{D2}R^3}{8}\int d\rho\,d\psi\,d\tau\,\sinh\rho
\sqrt{p^2+\frac{k}{2p^2}(k-2p)(1-\cos\alpha)}\,.
\ee
The equation of motion for $\alpha$ gives
\be
(k-2p)\sin\alpha=0\,,\qquad
\cS_\text{L.T, classical}=-p\,.
\ee
So either the solution has $p=k/2$ and arbitrary $\alpha$ or $\sin\alpha=0$ and 
$p$ is arbitrary. The first case is the solution 
presented before, while in the second it is not justified to 
use the D2-brane description, since it is singular, and a better description is 
in terms of $p$ fundamental strings.

Note that the two gauge choices for $C_1$ change the 
string charge by $k$, meaning that the charge is defined only modulo $k$. 
This is in agreement with the expectation from the gauge theory, where 
the $k$-th symmetric representation is analogous to the trivial one by the 
inclusion of an 't Hooft vertex.

It seems like the only regular configuration describes $k/2$ coincident 
Wilson loops (or a Wilson loop in the $k/2$ symmetric representation). 
We found singular solutions for other charges, but it is possible that our 
ansatz was too restrictive and that there are other regular solutions 
for arbitrary charges.
We note here that also in $AdS_5\times S^5$, while there are many explicit 
solutions for giant gravitons with fewer than 16  supercharges 
(see e.g. \cite{Mikhailov:2000ya}), only one class is known for $1/4$ BPS Wilson loops 
\cite{Drukker:2006zk}, so it is not too surprising if we cannot classify all possible 
D-branes dual to the $1/6$ BPS Wilson loops in the three-dimensional theory.

Furthermore note that usually the D-brane description of gauge theory operators 
is valid for representations of order $N$. The type IIA description is valid though 
for large $\lambda=N/k$, so a symmetric representation, whose dimension is 
capped by $k$, cannot approach $N$. 
This may explain why we find a regular solution only at the maximal value 
of $p$.

\subsection{D6-brane}
\label{sec-D6-adjoint}

The D2-brane solution seems to correspond to a Wilson loop in the symmetric 
representation, similar to the D3-brane in $AdS_5\times S^5$. There a Wilson loop 
in the anti-symmetric representation was described by a D5-brane, and the 
analog in our case is a D6-brane. We present the solution here.

This D6-brane will wrap a 5-dimensional submanifold of 
$\CP^3$, which we choose to 
have explicit $SU(2)\times SU(2)$ symmetry, as does the gauge theory 
operator.

Like the string and the D2-brane, the D6-brane will span an 
$AdS_2\subset AdS_4$. As usual, for the time-like 
Wilson line on antipodal points on $S^2$ it is parameterized by $\rho$ 
and $t$, while for the circular loop it is parameterized by $\rho$ and
$\psi$. Inside $\CP^3$ it will extend in the $\chi$, 
$\vartheta_1$, $\varphi_1$, $\vartheta_2$ and $\varphi_2$ directions 
at constant $\alpha$. We
also turn on an electric flux proportional to the volume form on $AdS_2$, 
so either $F=E\cosh\rho\,dt \wedge d\rho$, or $F=E\sinh\rho\,d\psi \wedge d\rho$.

The straight-line case will give a zero answer while the circle should give 
a non-trivial result. Due to that and the fact that the calculations are 
essentially identical, we write here the details for the case of the circle.

The action for this brane will include the DBI piece, as usual, and
the Wess-Zumino term coupling the pullback of $C_5$ \eqn{C5} 
to the world-volume field strength $F_{\psi\rho}=E\sinh\rho$
\be
\cS_\text{D6}=T_\text{D6}\int\Bigl[ e^{-\Phi}\sqrt{\det(g+2\pi\alpha'F)}
+2\pi i P[C_5]\wedge F\Bigr]\,.
\ee
Plugging in our ansatz we find
\be
\cS_\text{D6}=\frac{R^9T_\text{D6}}{2^{10}k^2}
\int %d\rho\,d\psi\,d\chi\,d\vartheta_1\,d\varphi_1\,d\vartheta_2\,d\varphi_2\,
\sinh\rho\sin\vartheta_1\sin\vartheta_2\left[
\sin^3\alpha\sqrt{1+\beta^2E^2}
-i\beta E\Big((\sin^2\alpha+2)\cos\alpha-2\Big)\right].
\ee
Here $\beta=8\pi k/R^3=\sqrt{2/\lambda}$ and $T_\text{D6}=1/(2\pi)^6$.

Integrating over the five remaining coordinates on $\CP^3$ 
gives a factor of $2^6\pi^3$ and we are left with an effective
theory on $AdS_2$. Now the 
equation of motion for $\alpha$ fixes the value of $E$
\be
i\beta E=-\cos\alpha\,.
\ee
The string charge carried by the D6-brane is the conjugate to the gauge field
\be
p=-i\frac{\delta S}{\delta E}
=\frac{\pi^3R^9T_\text{D6}\beta}{8k^2}(1-\cos\alpha)
=\frac{N}{2}(1-\cos\alpha)\,,
\ee
where we used that $N=R^6/(32\pi^2k)$. The value of $p$ ranges between
$0$ and $N$, where the appearance of $N$ is a
manifestation of the ``stringy exclusion principle,'' and is an
indication that this D-brane represents Wilson loops in anti-symmetric
representations.

Now we evaluate the classical action by performing a Legendre
transform, replacing the electric field with its conjugate $p$. We
also integrate over $AdS_2$ which gives a divergent answer, but whose
regularized area is $-2\pi$
\be
\cS_\text{L.T.}
=\cS-ipE
=-\frac{\pi^4R^9T_\text{D6}}{8k^2}\,\sin^2\alpha
=-\pi\sqrt{2\lambda}\,\frac{p\,(N-p)}{N}\,.
\label{action-D6}
\ee
This is indeed symmetric under $p\leftrightarrow N-p$, as should be
the case of the antisymmetric representation. Also for small $p$ it
agrees with the result of $p$ fundamental strings.

This construction is very similar to the D5-brane in $AdS_5\times S^5$ but 
some of the details are different. Here the relation between the charge $p$ 
and the angle $\alpha$ is trigonometric, while in the other case it is 
transcendental. Also the final answer \eqn{action-D6} is much simpler in 
this case. Note that the Gaussian matrix model reproduced the D5-brane 
result, so any modification of it to match the Wilson loop in $\cN=6$ CS 
should reproduce \eqn{action-D6}, once the relevant limit is taken (including 
non-planar corrections).

\subsection{Supersymmetry}

We turn now to checking the number of supersymmetries preserved by our 
string and D-brane solutions. We work in this section in 
Lorentzian signature and take the Wilson loop (and resulting $AdS_2$ 
surfaces) to be timelike.

As a first step one needs to choose a set of elfbeine and find the 
Killing spinors. This is done in Appendix~\ref{app-killing}, where the Killing spinors 
of M-theory on $AdS_4\times S^7$ in our coordinate system are found to be
\be
e^{\frac{\a}{4} ( \hat \g \g_4 - \g_{7\natural}   ) }
e^{\frac{\vartheta_1}{4} (  \hat \g \g_5 - \g_{8\natural}  ) }
e^{\frac{\vartheta_2}{4} ( \g_{79} + \g_{46} ) }
e^{-\frac{\xi_1}{2} \hat \g \g_\natural}
e^{-\frac{\xi_2}{2} \g_{58}}
e^{-\frac{\xi_3}{2} \g_{47}}
e^{-\frac{\xi_4}{2} \g_{69}}
e^{\frac{\rho}{2}\hat\gamma\gamma_1}
e^{\frac{t}{2}\hat\gamma\gamma_0}
e^{\frac{\theta}{2}\gamma_{12}}
e^{\frac{\phi}{2}\gamma_{23}}\epsilon_0
={\mathcal M}\epsilon_0\,,
\label{resks}
\ee
$\epsilon_0$ is a constant 32-component spinor and the Dirac matrices 
satisfy $\gamma_{012345678 9\natural}=1$.

The angles $\xi_i$ are the phases of $z_1,z_2,z_3,z_4$ from \eqn{zi} 
\be
\xi_1=\frac{2\varphi_1+\chi+\zeta}{4}\,,\qquad
\xi_2=\frac{-2\varphi_1+\chi+\zeta}{4}\,,\qquad
\xi_3=\frac{2\varphi_2-\chi+\zeta}{4}\,,\qquad
\xi_4=\frac{-2\varphi_2-\chi+\zeta}{4}\,.
\ee
Here $\zeta$ is the fiber direction on which the $\bZ_k$ orbifold acts.

To see which Killing spinors survive the orbifolding, we write the spinor 
$\epsilon_0$ in a basis which diagonalizes
\be
i\hat\gamma\gamma_\natural\epsilon_0=s_1\epsilon_0\,,\qquad
i\gamma_{58}\epsilon_0=s_2\epsilon_0\,,\qquad
i\gamma_{47}\epsilon_0=s_3\epsilon_0\,,\qquad
i\gamma_{69}\epsilon_0=s_4\epsilon_0\,.
\label{ss}
\ee
All the $s_i$ take values $\pm1$ and by our conventions on the product 
of all the Dirac matrices, the number of negative eigenvalues is even. 
Now consider a shift along the $\zeta$ circle, which changes all the 
angles by $\xi_i\to\xi_i+\delta/4$, the Killing spinors transform 
as
\be
{\mathcal M}\epsilon_0\to {\mathcal M} e^{i\frac{\delta}{8}
(s_1+s_2+s_3+s_4)}\epsilon_0\,.
\ee
This transformation is a symmetry of the Killing spinor when two of the $s_i$ eigenvalues 
are positive and two negative and not when they all have the same sign (unless $\delta$ 
is an integer multiple of $4\pi$). Note that on $S^7$ the radius of the $\zeta$ circle 
is $8\pi$, so the $\bZ_k$ orbifold of $S^7$ is given by taking $\delta=8\pi/k$. 
The allowed values of the $s_i$ are therefore
\be
(s_1,s_2,s_3,s_4)\in\left\{
\begin{matrix}
(+,+,-,-),\ (+,-,+,-),\ (+,-,-,+),\\
(-,+,+,-),\ (-,+,-,+),\ (-,-,+,+)
\end{matrix}
\right\}
\label{signs}
\ee
Each configuration represents four supercharges, so the orbifolding 
breaks $1/4$ of the supercharges (except for $k=1,2$) and leaves 
24 unbroken supersymmetries.

\subsubsection{Fundamental string}

Let us look now at the supersymmetries preserved by the string solution 
in Section~\ref{sec-F1}.
The string solution spans the $\rho$ and $t$ coordinates at some fixed values 
of the angles $\theta$ and $\phi$ (which we set to zero for simplicity). 

The supersymmetry condition for the fundamental string is\footnote{We denote
$\Gamma_\mu:=e_\mu^a\,\gamma_a$ where $e_\mu^a$ is the elfbein with
 $\mu$ a curved and $a$ a tangent space index.}
\be
(1-\Gamma)\, {\cal M}\, \epsilon_0=0 \qquad \text{with}\quad
\Gamma=\frac{1}{\mathcal L}\Gamma_{t\rho}\gamma_\natural
=\gamma_{01\natural}\,.
\ee

It is simple to rewrite the equation after multiplying from the left by $\cM^{-1}$. 
Setting $\alpha=0$ we have
\be
{\cal M}^{-1}\,\Gamma\, {\cal M}=
\Gamma\left(\cos^2\frac{\vartheta_1}{2}
+\cos\frac{\vartheta_1}{2}\sin\frac{\vartheta_1}{2}\,
e^{\xi_1\hat\gamma\gamma_\natural+\xi_2\gamma_{58}}
(\hat\gamma\gamma_5-\gamma_{8\natural})
-\sin^2\frac{\vartheta_1}{2}\hat\gamma\gamma_{58\natural}\right)
\ee
At the point $\vartheta_1=0$ we get the projector equation
\be
(1-\Gamma)\epsilon_0=0
\label{cond1}
\ee
Note that $\Gamma$ is an independent operator which commutes with 
$\gamma_{47}$, $\gamma_{58}$, $\gamma_{69}$ and $\hat\gamma\gamma_\natural$, 
so it will break half the supersymmetries. A localized string solution will therefore 
preserve 12 out of the 24 supercharges.

As mentioned above, we expect the dual of the $1/6$ BPS Wilson loop to be a 
string smeared along a $\CP^1$ subspace of $\CP^3$. This is given by taking 
arbitrary $\vartheta_1$, $\xi_1$ and $\xi_2$ at $\alpha=0$. We can see that if 
we impose the condition
\be
(\hat\gamma\gamma_5-\gamma_{8\natural})\epsilon_0=0\,,
\label{s1=s2}
\ee
then together with \eqn{cond1} this will guarantee that the equation 
$(1-\Gamma)\epsilon=0$ is satisfied at any point with $\alpha=0$.

The equation \eqn{s1=s2} is identical to the requirement $s_1=s_2$ 
\eqn{ss}. Therefore 
out of the six allowed eigenvalue combinations of the $s_i$ in \eqn{signs}
only two survive: $(+,+,-,-)$ and $(-,-,+,+)$. Together with \eqn{cond1} 
this means that four supercharges are preserved, as was found also in the 
gauge theory calculation.

\subsubsection{D2-brane}

The supersymmetries preserved by a D2-brane are determined by solving
the following equation on the D2-brane solution
\be\label{projD2}
\G \, \e = \e\,,
\ee
where $\e={\cal M}\,\e_0$ is the Killing spinor of the background, and where 
$\G$ for our D2-brane solution is given by (see e.g.~\cite{Cederwall:1996ri})
\be \Gamma=\frac{1}{\cL_{DBI}}\left(\Gamma^{(3)}
  +2\pi\alpha'F_{t\rho}\, \Gamma^{(1)}\gamma_\natural\right)\, .
\label{GammaD2}
\ee
Here
\be
\Gamma^{(3)}=\Gamma_{\mu_1\mu_2\mu_3}\, 
\frac{\partial x^{\mu_1}}{\partial \sigma^1}
\frac{\partial x^{\mu_2}}{\partial \sigma^2}\,
\frac{\partial x^{\mu_3}}{\partial \sigma^3}\,,
\ee
is the pullback of the curved space-time Dirac matrices in all 
world-volume directions and $\Gamma^{(1)}$ is the same, excluding the directions 
of the field strength $F_{t\rho}$. Plugging in our choice of coordinates and 
the details of the solution discussed in Section~\ref{sec-D2} we find
\bsp
&\G^{(3)} = -\frac{R^3}{16}\cosh \r \,\sin \a
\,\g_{017}\,,\\
&2\pi\alpha'F_{t\rho}\Gamma^{(1)} =  
\frac{R^3}{16}\cosh \r \,\sin\a\,\cos \a \, \g_{7}\,,\\
&{\cal L}_{DBI} = \frac{R^3}{16} \cosh \r \,\sin^2 \a\,.
\end{split}
\ee
And we therefore find that (\ref{projD2}) reads
\be\label{zwei}
\Bigl(\g_{01} + \cos \a \,\g_{\natural}\Bigr)\g_{7} \, \e = 
-\sin \a \, \e.
\ee
While we expect the D2-brane dual to the Wilson loop to be smeared 
over the directions parameterized by $\vt_1$, $\vp_1$, $\vt_2$ and $vp_2$, 
we start by considering a brane localized at the point where all these 
angles vanish. 
With $\vt_1=\vt_2=\vp_1=\vp_2=0$ the Killing spinor is greatly simplified
\be
\e |_{\vt_1=\vt_2=\vp_1=\vp_2=0} = 
e^{\frac{\a}{4} ( \hat \g \g_4 - \g_{7\natural}   ) }
e^{-\frac{\chi}{4}(s_1+s_2)} e^{\frac{\rho}{2}\hat\gamma\gamma_1}
e^{\frac{t}{2}\hat\gamma\gamma_0} \epsilon_0\,,
\ee
where we remind the reader that the $s_i$ are c-numbers obeying
$s_1+s_2+s_3+s_4=0$. We then rewrite (\ref{zwei}) in a suggestive
manner
\be
e^{\a\,\g_{7\natural}} \,\e = \g_{01\natural} \,\e\,.
\label{D2-proj}
\ee
Next we note that $ \hat \g \g_0$ and $\hat \g \g_1$
commute with both $\g_{7\natural}$ and $\g_{01\natural}$, and
therefore the $\rho$ and $t$ terms from the Killing spinor trivially
cancel.

Then, multiplying from the left by $e^{-\frac{\a}{4} ( \hat \g \g_4 - \g_{7\natural})}$ 
and commuting it though $\gamma_{01\natural}$ 
we find the following dependence on the angle $\alpha$
\be
e^{\a\,\g_{7\natural}}\,\e_0 
=  e^{-\frac{\a}{2}(\hat \g \g_4-\g_{7\natural})} \,\g_{01\natural} \,\e_0
\label{D2-proj-2}
\ee
It is now clear that in order to solve (\ref{zwei}) the
following two conditions must be imposed upon $\e_0$,
\be
\hat \g \g_4 \, \e_0 = - \g_{7\natural} \, \e_0, \qquad
\g_{01\natural} \, \e_0 = \e_0.
\ee 
Since $i\hat\g \g_\natural \, \e_0 = s_1 \, \e_0$ and $i \g_{47}
\, \e_0 = s_3 \, \e_0$, we see that the first of these two conditions
is that $s_1 = -s_3$, while the second condition, as we saw previously
for the fundamental string, acts independently to halve the
supersymmetries. Out of the six possible signs of the $s_i$ 
in \eqn{signs}, the condition $s_1=-s_3$ chooses four:
$(+,+,-,-)$, $(+,-,-,+)$,  $(-,+,+,-)$, and $(-,-,+,+)$. 
Recall that each choice corresponds to 4 supersymmetries, all of which
are halved by $\g_{01\natural}\,\e_0 = \e_0$. We have therefore a total of 8 out
of 24 supersymmetries preserved, i.e. the D2-brane at fixed
$\vt_1$, $\vt_2$, $\vp_1$ and $\vt_2$ is 1/3 BPS. 

A D2-brane localized at any other point will also preserve eight supercharges, 
we want to check which ones are shared by all of them. Consider then a 
D2-brane at the point $\vt_1=\pi$ and 
$\vt_2=\vp_1=\vp_2=0$. In this case the Killing spinor is
\be
\e |_{\vt_1=\pi,\ \vt_2=\vp_1=\vp_2=0} = 
e^{\frac{\a}{4} ( \hat \g \g_4 - \g_{7\natural}   ) }
e^{\frac{\pi}{4} (\hat \g \g_5 - \g_{8\natural})}
e^{-\frac{\chi}{4}(s_1+s_2)} e^{\frac{\rho}{2}\hat\gamma\gamma_1}
e^{\frac{t}{2}\hat\gamma\gamma_0} \epsilon_0\,.
\ee
Using relations like
\be
e^{-\frac{\pi}{4}\hat \g \g_5}
e^{\frac{\a}{4}  \hat \g \g_4 }
e^{\frac{\pi}{4}\hat \g \g_5}
=e^{-\frac{\a}{4} \g_{45} }\,,
\qquad\text{and}\qquad
e^{\frac{\pi}{4}\g_{8\natural}}
e^{-\frac{\a}{4}\g_{7\natural} }
e^{-\frac{\pi}{4}\g_{8\natural}}
=e^{-\frac{\a}{4}\g_{78} }\,,
\ee
transforms the projector equation to the form of \eqn{D2-proj-2} 
with the replacements $\hat\gamma\gamma_4\to-\gamma_{45}$, 
$\gamma_{7\natural}\to\gamma_{78}$, 
and $\gamma_{01\natural}\to s_1s_2\gamma_{01\natural}$ 
so the equation is solved for $\epsilon_0$ satisfying
\be
\g_{45} \, \e_0 = \g_{78} \, \e_0, \qquad
s_1s_2\g_{01\natural} \, \e_0 = \e_0.
\label{D2-pi}
\ee 
The first condition is analogous to imposing $s_2=-s_3$ and leaves the 
sign choices $(+,+,-,-)$, $(-,+,-,+)$,  $(+,-,+,-)$, and $(-,-,+,+)$. The second 
condition is a modification of the usual one 
$(\gamma_{01\natural}-1)\e_0=0$ for states with $s_1\neq s_2$.

Together with the previous condition, $s_1=-s_3$, for the D2-brane at 
$\vartheta_1=0$, this leaves only the two 
configurations $(+,+,-,-)$ and $(-,-,+,+)$. Now also $s_1=s_2$, 
so the second condition in \eqn{D2-pi} agrees with that in 
\eqn{D2-proj-2} giving a total of four real supercharges. These are the same 
supercharges preserved by the fundamental string after it was smeared 
on $\CP^1$.

A similar analysis can be done at any other value of the angles 
$\vt_1$, $\vt_2$, $\vp_1$ and $\vt_2$, but it is rather involved. 
A simpler route to the proof is to impose 
on the Killing spinor the conditions $s_1=s_2=-s_3=-s_4$ which eliminates 
from the Killing spinor all dependence on these angles
\be
\e=
e^{-\frac{\a}{2} \g_{7\natural} }
e^{-\frac{\chi}{2}s_1}e^{\frac{\rho}{2}\hat\gamma\gamma_1}
e^{\frac{t}{2}\hat\gamma\gamma_0} \epsilon_0\,.
\label{KS2}
\ee
Commuting the Dirac matrices in the projector equation \eqn{D2-proj} 
through we find that after imposing $(\g_{01\natural} -1)\e_0=0$, 
the projector equation is satisfied.

We conclude that after smearing the D2-brane, we end up with a configuration 
which is $1/6$-BPS, like the Wilson loop operators in the gauge theory.

\subsubsection{D6-brane}

The supersymmetry projector associated to the D6-brane is $\G \, \e =
\e$, where now (see e.g.~\cite{Cederwall:1996ri})
\be
\Gamma=\frac{1}{\cL_{DBI}}\left(\Gamma^{(7)}
+2\pi\alpha'\, F_{t\rho}\,\Gamma^{(5)}\gamma_\natural\right)\,,
\label{GammaD6}
\qquad
\Gamma^{(7)}=\Gamma_{\mu_1\ldots\mu_7}\, \frac{\partial x^{\mu_1}}{\partial \sigma^1}
\, \ldots\, \frac{\partial x^{\mu_7}}{\partial \sigma^7}\,.
\ee
$\Gamma^{(5)}$ again is the same as $\Gamma^{(7)}$, excluding the directions 
of the field strength $ F_{t\rho}$. Plugging in our choice of coordinates and 
the details of the solution presented in Section~\ref{sec-D6-adjoint}, we find
\be
\Bigl( \g_{01} + \cos \a \,\g_\natural \Bigr) \g_{56789} \, \e = \sin
\a \, \e.
\label{eins}
\ee
The form of this projector is quite easy to understand. $\G^{(7)}$, 
$\G^{(5)}$ and the Lagrangian share the same volume element on 
$\CP^3$ and with the field-strength also that of $AdS_2$. Then the 
remaining factors come from $\beta E=-\cos\alpha$ and a factor 
of $\sqrt{1-\b^2 E^2}=\sin\alpha$ in the DBI Lagrangian.

The equation is very similar to that in the D2-brane case. It needs to be 
checked for all values of $\vt_1$, $\vp_1$, $\vt_2$, $\vp_2$ and $\chi$. 
One first chooses a pair of points and verifies that the same conditions 
as for the fundamental string and the D2-brane are necessary at those 
two points. Then we can use these conditions, in particular
$s_1=s_2=-s_3=-s_4=\pm1$ to express the Killing spinor as \eqn{KS2}
\be
\e=
e^{-\frac{\a}{2} \g_{7\natural} }
e^{-\frac{\chi}{2}s_1}e^{\frac{\rho}{2}\hat\gamma\gamma_1}
e^{\frac{t}{2}\hat\gamma\gamma_0} \epsilon_0\,.
\ee
Now we rewrite (\ref{eins}) as
\be
e^{-\a\,\g_{56789\natural}} \,\e 
=e^{\a\,\g_{7\natural}} \,\e 
= \g_{01\natural} \,\e\,.
\ee
Since $ \hat \g \g_0$ and $\hat \g \g_1$
commute with $\g_{7\natural}$, $\g_{69}$, $\g_{58}$, and
$\g_{01\natural}$, the $\rho$ and $t$ terms from the
Killing spinor trivially cancel. We are left with
\be
e^{\frac{\a}{2}\g_{7\natural}}\,\e_0 
= \g_{01\natural} \, e^{-\frac{\a}{2} \g_{7\natural}} \,\e_0
\ee
The alpha dependence drops since $\gamma_{01\natural}$ 
anti-commutes with $\gamma_{7\natural}$, so finally we are left 
with the condtion $\g_{01\natural}\, \e_0 = \e_0$. We
have therefore found that the D6-brane preserves the same supersymmetries
of the smeared fundamental string the D2-branes and of the Wilson loop operator.

\section{Discussion}
\label{sec-discussion}

In this paper we studied supersymmetric Wilson loops in the $\cN=6$ 
Chern-Simons theory constructed by Aharony et al. \cite{Aharony:2008ug}. 
The natural Wilson loop observable couples to a bi-linear of the scalar fields 
and we studied the simplest such loops, with the geometry of a line or a 
circle both in the gauge theory (at order $\lambda^2$) and at strong coupling 
using fundamental strings, D2-branes and D6-branes in $AdS_4\times\CP^3$.

In the maximally supersymmetric theory in four dimensions the circular 
Wilson loop has a non-trivial expression which can be matched between 
the gauge theory and string theory by an exact interpolating function. 
It would be interesting to see if such results apply also here, though 
the basic Wilson loop observable preserves $1/6$ of the supercharges, not 
$1/2$.

It is a rather puzzling fact that the natural Wilson loops preserve only four 
supercharges. A fundamental string ending along a straight line on the boundary 
of $AdS_4$ and localized on $\CP^3$ preserves 12 supercharges. In order 
to match with the gauge theory observable and its $SU(2)\times SU(2)$ 
symmetry we smeared the string over a $\CP^1$, and it indeed
broke the supersymmetry down to $1/6$. But the question remains what is the 
gauge theory dual to a {\em localized} fundamental string.

We comment below on some possibilities to construct such operators, but will 
not pursue them here further.

The Wilson loop \eqn{adjoint-WL} preserves 4 supercharges which match 
with 4 out of the 12 supercharges preserved by the localized fundamental 
string, but it breaks the other 8. Those other eight supercharges will not 
annihilate this loop, but transform it into a different loop and by repeated action 
one can generate a full multiplet of Wilson loops. This multiplet is closed 
under the action of all the required 12 supercharges, so the state created 
by integrating over all those Wilson loops with flat measure will necessarily 
preserve these 12 supercharges.

This is a standard way of enhancing symmetry, by integrating over the 
zero modes of the broken symmetry. It is guaranteed to give an object with 
at least the desired symmetry, but it might also lead to a trivial operator, the 
identity or 0. It would be interesting to construct this operator explicitly 
and study its properties.

Let us point out here another possible construction of a supersymmetric 
Wilson loop. Consider the purely bosonic operator \eqn{bosonic-WL} 
with the holonomy in {\em both} of the gauge groups and in opposite 
representations. Such a Wilson loop may be writen schematically as
\be
W=\Tr\cP\exp \left(i\int(A_\mu-\hat A_\mu)dx^\mu\right)\,.
\label{bi-WL}
\ee
The relative sign was put in by hand to represent the fact that if the first 
group is in the fundamental representation the second one is in the 
anti-fundamental, and hence the gauge fields act on the fields from the 
right, rather than the left.

A naive tree-level calculation of the supersymmetry variation of this loop will show 
that it is invariant under {\em all} the supersymmetries, the variation of 
$\hat A$ canceling that of $A$ after taking the trace. We do not expect this 
to extend beyond the tree level, and indeed the expectation value of this loop 
will suffer from divergences at order $\lambda^2$. But it is possible that 
this loop will become supersymmetric once augmented with the correct 
scalar insertions.

One can also use this operator to construct open Wilson loops, by putting 
a bi-fundamental field, say $C_I$ at one end and an adjoint field 
$\bar C^J$ at the other. Furthermore, one can start the open Wilson-loop 
at one $C_I$ and then continue to insert more $C_I$ fields along its path. 
After each scalar field the representation of the Wilson loop will change 
(to a product representation of the fundamental-antifundamental). Because 
the operator $(C_I)^k$ is gauge invariant (with the inclusion of an 't Hooft 
vertex), after $k$ insertions, the Wilson loop can end.

These are two ways of constructing open Wilson loops. While we don't expect 
a loop of finite extent to be supersymmetric, one can consider the infinite line 
with a distribution of bi-fundamental scalar field insertions. With an appropriate 
choice of scalars (the simplest being all identical), the same naive argument 
would lead one to conclude that this Wilson loop preserves some 
supersymmetries. We leave a closer examination of those Wilson loops to the 
future.

\subsection*{Acknowledgments}
We would like to thank Ofer Aharony, Fernando Alday, Abhishek Agarwal,
Adi Armoni, Jaume Gomis, Rajesh Gopakumar, Johannes Henn, Shiraz Minwalla,
Constantinos Papageorgakis, Spenta Wadia, Konstantin Wiegandt, Xi Yin and all the
participants of the Monsoon Workshop for stimulating discussion.  N.D.
acknowledges the welcome hospitality of the Tata Institute for
Fundamental Research and the ICTS, Mumbai during the course of this
work.  D.Y. acknowledges the support of the National Sciences and
Engineering Research Council of Canada (NSERC) in the form of a
Postdoctoral Fellowship.  This work was supported by the Volkswagen
Foundation.

%\pagebreak

\appendix
\section{${\cal N}=6$, $d=3$ super Cherns-Simons-matter action and Feynman rules}
\label{appA}

Here we summarize the action and conventions for the perturbative computations.
The field content consists of two $U(N)$ gauge fields $(A_\mu)_{ij}$ and
$(\hA_\mu)_{\hi\hj}$, the complex fields $(C_I)_{i\hi}$ and $({\bar C}^{I})_{\hi i}$ as
well as the fermions $(\psi_I)_{\hi i}$ and $({\bar\psi}^{I})_{i \hi }$ in 
the $({\bf N},{\bf \bar N})$ and $({\bf \bar N},{\bf N})$ of $U(N)$ respectively,
$I=1,2,3,4$ is the $SU(4)_R$ index. We employ the covariant gauge fixing function 
$\partial_\mu A^\mu$ for both gauge fields and have two sets of ghosts $(\bar c,c)$ and
$(\bar{\hat c},\hat c)$. We work with the Euclidian space action 
(see \cite{Chen:1992ee,Aharony:2008ug,Benna:2008zy})
\begin{align}
S_\text{CS} & = -i\frac{ k}{4\pi}\,\int d^3x\, \varepsilon^{\mu\nu\rho} \,\Bigl [\,
\Tr (A_\mu\partial_\nu A_\rho+\frac{2}{3}\,A_\mu A_\nu A_\rho)- 
\Tr (\hA_\mu\partial_\nu \hA_\rho+\frac{2}{3}\, \hA_\mu \hA_\nu \hA_\rho)\, \Bigr ] \nn\\
S_\text{gf} & = \frac{k}{4\pi}\, \int d^3 x\, \Bigl [\,\frac{1}{\xi}\, \Tr(\partial_\mu A^\mu)^2
+\Tr(\partial_\mu \bar c\, D_\mu c) - \frac{1}{\xi}\, \Tr(\partial_\mu \hA^\mu)^2
+\Tr(\partial_\mu \bar{ \hat c}\, D_\mu \hat c) \, \Bigr ]\nn\\
S_\text{Matter} & = \int d^3 x\, \Bigl [\, \Tr(D_\mu\, C_I\, D^\mu {\bar C}^{I}) + i\, \Tr(\bar\psi^I
\, \slsh{D}\, \psi_I)\, \Bigr ] + S_\text{int}
\end{align}
Here $S_\text{int}$ are the sextic scalar potential and $\psi^2 C^2$ Yukawa type potentials
spelled out in \cite{Aharony:2008ug}. The matter covariant derivatives are defined as
\be
\begin{aligned}
D_\mu C_I &= \partial_\mu C_I + i (A_\mu\, C_I - C_I\, \hA_\mu) \\ 
D_\mu {\bar C}^{I} &= \partial_\mu {\bar C}^{I} - i ({\bar C}^{I}\,A_\mu -  \hA_\mu\, {\bar C}_{I})  \\ 
D_\mu \psi_I &= \partial_\mu \psi_I + i (\hA_\mu\, \psi_I - \psi_I\, A_\mu)  \\ 
D_\mu {\bar\psi}^{I} &= \partial_\mu {\bar \psi}^{I} - i ({\bar \psi}^{I}\,\hA_\mu -  A_\mu\, 
{\bar \psi}^{I})\, . 
\end{aligned}
\label{cov-deriv}
\ee
From this we read off the momentum space propagators
\be
\begin{aligned}
\langle (A_\mu)_{ij}(p)\, (A_\nu)_{kl}(-p) \rangle_0 & = 
\frac{2\pi}{k}\, \delta_{il}\, \delta_{jk}\, \Bigr [ \varepsilon_{\mu\nu\rho}\, p^\rho +
\xi\, \frac{p_\mu p_\nu}{p^2}\, \Bigr ]\, \frac{1}{p^2}  \\
\langle (\hA_\mu)_{ij}(p)\, (\hA_\mu)_{kl}(-p) \rangle_0 & = 
-\frac{2\pi}{k}\, \delta_{il}\, \delta_{jk}\, \Bigr [ \varepsilon_{\mu\nu\rho}\, p^\rho +
\xi\, \frac{p_\mu p_\nu}{p^2}\, \Bigr ]\, \frac{1}{p^2}  %\\
\end{aligned}
\ee
\be
\begin{aligned}
\langle (c)_{ij}(p)\, (\bar c)_{kl}(-p) \rangle_0 & = 
\frac{2\pi}{k}\,  \delta_{il}\, \delta_{jk}\, \frac{1}{p^2} \\
\langle (\hat c)_{ij}(p)\, (\bar {\hat c})_{kl}(-p) \rangle_0 & = -\frac{2\pi}{k}\,  \delta_{il}\, \delta_{jk}\, 
\frac{1}{p^2} %\\
\end{aligned}
\ee
\be
\begin{aligned}
\langle (C_I)_{i\hi}(p)\, ({\bar C}^{J})_{\hj j}(-p) \rangle_0 &=  \delta_{I}^J\,
\delta_{ij}\, \delta_{\hi \hj}\,
\frac{1}{p^2} \\
\langle (\psi_I)_{\hi i}(p)\, ({\bar\psi}^{J})_{j \hj }(-p) \rangle_0 &= -\delta_{I}^J\,
 \delta_{ij}\, \delta_{\hi \hj}\,\frac{1}{p\!\! /} 
\end{aligned}
\ee
We also note the relevant Fourier transformations to configuration space:
\be
\left[ \frac{\delta_{\mu\nu}}{p} - \frac{p_\mu p_\nu}{p^3}\right]_{d=3} \to
\frac{\delta_{\mu\nu}}{2\pi^2\, x^2} - \frac{1}{4\pi^2}\partial_\mu\partial_\nu \log x^2\, , \qquad
\left[ \frac{1}{p^2} \right]_{d=3} \to \frac{1}{4\pi}\, \frac{1}{x}
\ee

\subsection{The gluon self-energy}

The one-loop correction to the gluon self energy from gluon and ghost contributions is known to
vanish (see e.g.~\cite{Chen:1992ee}). We here evaluate the matter contribution.

For bosons in the loop there are two graphs to consider, the four-valent bubble vanishes in dimensional
regularization. The other graph comes from expanding the cubic vertex to second order
from $e^{-S_\text{Matter}}$:
\begin{align}
\Bigl \langle ( \, i\,  
\Tr(A_\mu\, C_I\,\partial_\mu\bar C^I\,-  \p_\m C_I \barC^I A_\m ))\, 
 ( \, i\,  \Tr(A_\mu\, C_I\,
\partial_\mu\bar C^I\,-  \p_\m C_I \bar C^I A_\m ))
\Bigr \rangle
\end{align}
Contracting, Fourier transforming and amputating the gluon legs yields the 
self-energy contribution
\begin{equation}
\label{PiB}
\Pi^{(B)}_{\mu\nu}(p)=   N\, \delta^I_I\, \int \frac{d^3 k}{(2\pi)^3}\,
\frac{(2k+p)_\mu (2k+p)_\nu}{k^2\, (p+k)^2} .
\end{equation}
This is to be contracted with two gluon propagators (we use Landau gauge $\xi=0$ from
now on) to get the one-loop corrected gluon propagator 
\begin{equation}
G^{(B,1)}_{\mu\nu}(p) = \left(\frac{2\pi}{k}\right)^2\, 
\frac{\varepsilon_{\mu\rho\kappa}p^\kappa}{p^2}\, \Pi^{(B)}_{\rho\lambda}(p)\,
\frac{\varepsilon_{\lambda\nu\delta}p^\delta}{p^2}\, ,
\end{equation}
In this expression we see that the term in the integral proportional to $p_\nu$ 
in \eqn{PiB} drops out.
Performing the integral in \eqn{PiB} in dimensional regularization yields a finite
result. Contracting with the two $\varepsilon$-tensors we find
\begin{equation}
G^{(B,1)}_{\mu\nu}(p) = \left(\frac{2\pi}{k}\right)^2\,  \frac{N\, \delta^I_I}{16}\,
\frac{1}{p}\, \left (\, \delta_{\mu\nu} - \frac{p_\mu p_\nu}{p^2}\, \right )\, .
\end{equation}

Turning to the fermionic contributions to the loop  we need to contract
\begin{equation}
\Bigl \langle ( \, i\,  \Tr(\bar \psi^I\, i\slsh{A}\, \psi_I\, )\, 
( \, i\,  \Tr(\bar \psi^I\, i\slsh{A}\, \psi_I\, )\, \Bigr \rangle
\end{equation}
yielding
\begin{equation}
\label{PiF}
\Pi^{(F)}_{\mu\nu}(p)= - N\, \delta^I_I\, \int \frac{d^3k}{(2\pi^3)}\,
\frac{\text{tr}(\gamma_\mu\, (\pslash+\kslash)\, \gamma_\nu\, \kslash)}{k^2\, (p+k)^2}.
\end{equation}
We note
\begin{equation}
\text{tr}(\gamma_\mu \pslash \gamma_\nu\kslash) = 2 \, (-\delta_{\mu\nu}\, p\cdot k
+k_\mu p_\nu + p_\mu k_\nu)\, ,
\end{equation}
where the last two terms vanish upon contraction with the epsilon
tensors of the attached gluon propagators. This leaves the $p\cdot k$ term. Upon using a Feynman
parameter $\a$, only the momentum shift of $k \rightarrow k - (1-\a)p$
will survive integration. This gives
\be
2\, N \d^I_I \d_{\m\n} \int \frac{d^3 k}{(2\pi)^3}
\int_0^1 d\a \frac{-p^2(1-\a)}{[k^2 + \a(1-\a)p^2]^2}
= -2\,N \d^I_I \d_{\m\n} \frac{p \pi^{3/2}\G(1/2)}{(2\pi)^3} \int_0^1 d\a
\sqrt{\frac{1-\a}{\a}} = \frac{ -N \d^I_I \d_{\m\n}p}{8}
\ee
We must also consider the term:
\begin{equation}
\text{tr}(\gamma_\mu \kslash \gamma_\nu\kslash) = 2 \, (-\delta_{\mu\nu}\, k\cdot k
+2k_\mu k_\nu)\, .
\end{equation}
Upon shifting the momentum $k$ as above we obtain
\be
-2 \d_{\m\n} \left[ k^2 + (1-\a)^2p^2\right] + 4\left[k_\m k_\n + (1-\a)^2
  p_\m p_\n \right] \rightarrow -2\d_{\m\n} \left[ \frac{k^2}{3}
  +(1-\a)^2 p^2 \right]
\ee
where we have symmetrized the $k_\m k_\n$ integral, and removed the
$p_\m p_\n$ term as it is killed by epsilon contractions. Integrating over
$k$ we find
\be
-2N \d^I_I\frac{ \d_{\m\n}}{(2\pi)^3} \int_0^1 d\a \Bigl[ 
-\frac{1}{3} \d_{\m\n} \frac{3}{2} \pi^{3/2} \G(-1/2) \sqrt{\a(1-\a)}
p
- \d_{\m\n} \frac{(1-\a)^{3/2}}{\sqrt{\a}} \pi^{3/2}\G(1/2)p \Bigr]
= \frac{N\d^I_I \d_{\m\n} p}{16}
\ee

We therefore have
\be
G^{(F,1)}_{\mu\nu}(p) = \left(\frac{2\pi}{k}\right)^2\,  \frac{N\, \delta^a_a}{16}\,
\frac{1}{p}\, \left (\, \delta_{\mu\nu} - \frac{p_\mu p_\nu}{p^2}\, \right )\, ,
\ee
and so the combined bosonic and fermionic matter contributions yield
\be
G^{(1)}_{\mu\nu}(p) = G^{(B,1)}_{\mu\nu}(p) +G^{(F,1)}_{\mu\nu}(p) = 
 \left(\frac{2\pi}{k}\right)^2\,  \frac{1}{8}\,
\frac{N\, \delta^a_a}{p}\, \left (\, \delta_{\mu\nu} - \frac{p_\mu p_\nu}{p^2}\, \right )\, .
\ee

\section{Killing spinors}
\label{app-killing}

In this appendix we derive an explicit form for the Killing spinors in the 
coordinate system where the metric on $AdS_4$ is \eqn{AdS-metric}
\be
ds_{AdS_4}^2=R^2\left[d\rho^2-\cosh^2\rho\,dt^2
+\sinh^2\rho\big(d\theta^2+\sin^2\theta\,d\phi^2\big)\right]\,.
%\label{global-ads}
\ee
and the metric on $S^7$ is given by \eqn{S7-metric}
\be
\begin{aligned}
ds^2_{S^7}=\frac{R^2}{4}\Bigg[&
d\alpha^2
+\cos^2\frac{\alpha}{2}(d\vartheta_1^2+\sin^2\vartheta_1^2\,d\varphi_1^2)
+\sin^2\frac{\alpha}{2}(d\vartheta_2^2+\sin^2\vartheta_2^2\,d\varphi_2^2)
\\&
+\sin^2\frac{\alpha}{2}\cos^2\frac{\alpha}{2}
(d\chi+\cos\vartheta_1\,d\varphi_1-\cos\vartheta_2\,d\varphi_2)^2
\\&
+\left(\frac{d\zeta}{2}+\cos^2\frac{\alpha}{2}\cos\vartheta_1\,d\varphi_1
+\sin^2\frac{\alpha}{2}\cos\vartheta_2\,d\varphi_2+\frac{\cos\alpha}{2}\,d\chi\right)^2
\Bigg].
\end{aligned}
\ee
We take the elfbeine to be
\be
\begin{gathered}
e^0=\frac{R}{2}\cosh\rho\,dt\,,\quad
e^1=\frac{R}{2}\,d\rho\,,\quad
e^2=\frac{R}{2}\sinh\rho\,d\theta\,,\quad
e^3=\frac{R}{2}\sinh\rho\sin\theta\,d\phi\,,\\
e^4 =\frac{R}{2}  d\a, \qquad 
e^5 =\frac{R}{2}  \cos\frac{\a}{2}\,d\vartheta_1,\qquad 
e^6 =\frac{R}{2}  \sin\frac{\a}{2} \, d\vartheta_2,\\
e^7 = \frac{R}{2}  \cos\frac{\a}{2}\sin\frac{\a}{2} \Bigl( \cos \vartheta_1 \,
d\varphi_1 - \cos\vartheta_2\,d\varphi_2 + d\chi \Bigr),\\
e^8 = \frac{R}{2}  \cos\frac{\a}{2} \sin\vartheta_1\,d\varphi_1,\quad
e^9 =\frac{R}{2}  \sin\frac{\a}{2}\sin\vartheta_2 \, d\varphi_2\,,\\
e^\natural=-\frac{R}{4}\left(d\zeta+2\cos^2\frac{\alpha}{2}\cos\vartheta_1\,d\varphi_1
+2\sin^2\frac{\alpha}{2}\cos\vartheta_2\,d\varphi_2+\cos\alpha\,d\chi\right).
\end{gathered}
\ee

To find the relevant Killing spinor equation for this background we look at 
the supersymmetry transformation of the gravitino
\be
\delta\Psi_\mu=D_\mu\epsilon-\frac{1}{288}\Big(
\Gamma_\mu^{\,\n\l\r\s}
-8\delta_\mu^\nu\Gamma^{\lambda\r\sigma}\Big)F_{\n\l\r\s}\epsilon\,,\qquad
D_\mu\epsilon=\partial_\mu\epsilon+\frac{1}{4}\omega_\mu^{ab}\gamma_{ab}\epsilon\,.
\ee
The 4-form corresponding to the $AdS_4 \times S^7$ solution is
$F_{\nu\lambda\rho\sigma} = 6\, \varepsilon_{\nu\lambda\rho\sigma}$, 
where the epsilon symbol is the volume form on $AdS_4$ (so the indices take 
the values $0,1,2,3$). Plugging this into the variation
above one finds the Killing spinor equation
\be
D_\m \e = \frac{1}{2} \hat \g \g_\m \e
\ee
where $\m$ runs over all 11 coordinates, and $\hat \g =\g^{0123}$. Note that 
small $\gamma$ have tangent-space indices while capital $\Gamma$ carry 
curved-space indices. Calculating the spin-connection for 
our chosen elfbeine, we find the following explicit Killing spinor equations
\be
\begin{aligned}
\partial_t\epsilon&=\frac{1}{2}\hat\gamma\gamma_1\,e^{\rho\,\hat\gamma\gamma_0}\epsilon\,,\\
\partial_\rho\epsilon&=\frac{1}{2}\hat\gamma\gamma_1\epsilon\,,\\
\partial_\theta\epsilon&=\frac{1}{2}\gamma_{12}\,e^{-\rho\,\hat\gamma\gamma_1}\epsilon\,.\\
\partial_\phi\epsilon&=\frac{1}{2}\left(\coth\rho\,\gamma_{13}+\cos\theta\,\gamma_{23}
+\sinh\rho\sin\theta\,\hat\gamma\gamma_3\right)\epsilon=0\,,\\
\end{aligned}
\ee
and
\be
\begin{aligned}
\partial_\alpha\epsilon=&\frac{1}{4}\,( \hat\gamma\gamma_4  -\g_{7\natural} )\epsilon\,,\\
\partial_{\vartheta_1}\epsilon=&
\frac{1}{4}\left(\hat\gamma\gamma_5\,e^{-\frac{\alpha}{2}\hat\gamma\gamma_4}
- \g_{8\natural}\,e^{\frac{\alpha}{2}\gamma_{7\natural}}\right)\epsilon\,,\\
\partial_{\vartheta_2}\epsilon=&
\frac{1}{4}\left(\gamma_{46}\,e^{-\frac{\alpha}{2}\hat\gamma\gamma_4}
+\gamma_{79}\,e^{\frac{\alpha}{2}\gamma_{7\natural}}\right)\epsilon\,,\\
\partial_{\varphi_1}\epsilon=&
\frac{1}{4}\left(\cos\vartheta_1\,\gamma_{58}
-\cos\vartheta_1\,\hat\gamma\gamma_\natural\,
e^{\frac{\alpha}{2}(\gamma_{7\natural}-\hat\gamma\gamma_4)}
+\sin\vartheta_1\left(\hat\gamma\gamma_8\,e^{-\frac{\alpha}{2}\hat\gamma\gamma_4}
+\gamma_{5\natural}\,e^{\frac{\alpha}{2}\gamma_{7\natural}}\right)\right)\epsilon\,,\\
\partial_{\varphi_2}\epsilon=&
\frac{1}{4}\left(\cos\vartheta_2\,\gamma_{69}
-\cos\vartheta_2\,\gamma_{47}\,
e^{\frac{\alpha}{2}(\gamma_{7\natural}-\hat\gamma\gamma_4)}
+\sin\vartheta_2\left(\gamma_{49}\,e^{-\frac{\alpha}{2}\hat\gamma\gamma_4}
+\gamma_{67}\,e^{\frac{\alpha}{2}\gamma_{7\natural}}\right)\right)\epsilon\,,\\
\partial_\chi\epsilon=
&\frac{1}{8}\left((\g_{47} -\hat \g \g_{\natural})e^{-\alpha\hat\gamma\gamma_4}
 + \g_{69} - \g_{58}\right)\e\,,\\
\partial_\zeta\epsilon=&
-\frac{1}{8}(\g_{58} + \g_{69} + \g_{47}+ \hat\gamma\gamma_\natural)\epsilon\,.\\
\end{aligned}
\ee
These equations are solved by the following Killing spinor
\be
e^{\frac{\a}{4} ( \hat \g \g_4 - \g_{7\natural}   ) }
e^{\frac{\vartheta_1}{4} (  \hat \g \g_5 - \g_{8\natural}  ) }
e^{\frac{\vartheta_2}{4} ( \g_{79} + \g_{46} ) }
e^{-\frac{\xi_1}{2} \hat \g \g_\natural}
e^{-\frac{\xi_2}{2} \g_{58}}
e^{-\frac{\xi_3}{2} \g_{47}}
e^{-\frac{\xi_4}{2} \g_{69}}
e^{\frac{\rho}{2}\hat\gamma\gamma_1}
e^{\frac{t}{2}\hat\gamma\gamma_0}
e^{\frac{\theta}{2}\gamma_{12}}
e^{\frac{\phi}{2}\gamma_{23}}\epsilon_0
={\mathcal M}\epsilon_0\,,
\label{Killing}
\ee
where the $\xi_i$ are given by
\be
\xi_1=\frac{2\varphi_1+\chi+\zeta}{4}\,,\qquad
\xi_2=\frac{-2\varphi_1+\chi+\zeta}{4}\,,\qquad
\xi_3=\frac{2\varphi_2-\chi+\zeta}{4}\,,\qquad
\xi_4=\frac{-2\varphi_2-\chi+\zeta}{4}\,.
\ee
In \eqn{Killing} $\epsilon_0$ is a constant 32-component spinor 
and the Dirac matrices were chosen such that
$\gamma_{012345678 9\natural}=1$. 
A similar calculation in a different coordinate system was done in 
\cite{Nishioka:2008ib}.

One may also consider the $AdS_4$ in terms of $AdS_2$ slices
\eqn{AdS-metric-2}
\be
ds^2 = du^2 + \cosh^2 u \Bigl( -\cosh^2 \r \, dt^2 + d\r^2 \Bigr) +
\sinh^2 u \, d\phi^2,
\ee
a vierbein basis being given by
\be
e^0 = \frac{R}{2} \cosh u \cosh \r \, dt, \quad e^1 = \frac{R}{2}
\cosh u \, d\r, \quad
e^2 = \frac{R}{2} du , \quad e^3 = \frac{R}{2} \sinh u \, d\phi,
\ee
leading to the following spin connection
\be
\o^{01} = \sinh \r \, dt, \quad \o^{02} =  \sinh u\cosh \r \, dt,\quad
\o^{12} = \sinh u \, d\r, \quad
\o^{23} = - \cosh u \, d\phi.
\ee
In these coordinates the final four factors in the Killing spinor in
(\ref{Killing}) are replaced by
\be\label{altAdS4KS}
e^{\frac{u}{2} \hat \g \g_2}
e^{\frac{\phi}{2} \g_{23}}
e^{\frac{\r}{2} \hat \g \g_1}
e^{\frac{t}{2} \hat \g \g_0}.
\ee

\end{fmffile}

\pagebreak

\end{document}